\documentclass[twocolumn,showpacs,preprintnumbers,amsmath,amssymb,letter]{revtex4}
\PassOptionsToPackage{caption=false}{subfig}
\usepackage{graphicx}
\usepackage{dcolumn}
\usepackage{bm}
\usepackage{subfig}

\begin{document}

\title{Continuum simulations of shocks and patterns in vertically oscillated 
granular layers}

\author{J. Bougie, K. Duckert}
 \affiliation{Department of Physics,
{Loyola University Chicago, Chicago, IL 60660}}

\date{September 14, 2010}

\begin{abstract}
We study interactions between 
shocks and standing-wave patterns in vertically oscillated
layers of granular media using three-dimensional, 
time-dependent numerical solutions
of continuum equations to Navier-Stokes order.
We simulate a layer of grains atop a plate that oscillates
sinusoidally in the direction of gravity. 
Standing waves form stripe patterns when the
accelerational amplitude of the plate's oscillation exceeds
a critical value.  
Shocks also form with each collision between the layer and the plate;
we show
that pressure gradients formed by these shocks
cause the flow to reverse direction within the layer.
This reversal leads to an oscillatory state of the pattern
that is subharmonic with
respect to the plate's oscillation.
Finally, we study the relationship between shocks and patterns
in layers oscillated at various frequencies and show that
the pattern wavelength increases monotonically as
the shock strength increases.

\end{abstract}


\maketitle

\section{Introduction}

\subsection{Background}

Vertically oscillated granular layers provide an 
important testbed for 
granular research. When shaken vertically,
flat layers of grains exhibit 
convection \cite{knight96}, clustering \cite{falcon99}, 
shocks \cite{goldshtein2}, 
steady-state flow fields far from the plate \cite{brey01}, 
floating particle clusters \cite{eshuis},
and standing-wave pattern formation \cite{melo}.

In this paper, we model granular media by numerically solving
a set of time-dependent continuum equations
for the rapid flow of dissipative particles in three-dimensions.  
We use these simulations to investigate
the relationship between shocks and standing-wave patterns in vertically
oscillated granular layers.

\subsection{Granular hydrodynamics}

A successful theory of granular hydrodynamics 
would allow scientists and 
engineers to apply the powerful methods of fluid 
dynamics to granular flow.  
Despite experimental \cite{bocquet, rericha, eshuis2010} and computational 
\cite{ramirez,carrillo2008, bougie2005, bougie2010, eshuis2010} evidence
demonstrating the potential utility of hydrodynamics models for grains,
a general set of hydrodynamic governing equations is not yet recognized
for granular media \cite{dufty2002,campbell,tsimring, rericha2004}.

Several proposed rapid granular flow models use equations of motion
for continuum fields: number density $n$, velocity ${\bf u}$, and
granular temperature $T$ ($\frac{3}{2} T$ 
is the average kinetic energy due
to random particle motion)  \cite{haff,jenkinsandsavage, lun1984}.  
In one approach, particle interactions are 
modeled with binary, inelastic
hard-sphere collision operators in kinetic theory to derive continuum
equations to Euler \cite{goldshtein1}, Navier-Stokes
\cite{jenkinsandrichman}, and Burnett 
\cite{selaandgoldhirsch} order.  
We use three-dimensional (3D) simulations of continuum equations to
Navier-Stokes order to investigate shocks and standing-wave patterns in
shaken granular layers.

\subsection{Standing wave patterns in oscillated granular layers}

A granular layer of depth $H$ atop a plate that is oscillated 
sinusoidally in the direction of gravity with 
frequency $f$ and amplitude
$A$ will leave the plate at some time during the 
oscillation cycle if the maximum 
acceleration of the plate $a_{\rm max}=A\left(2 \pi f \right)^2$ 
is greater than
the acceleration of gravity $g$.  In other words, 
the layer leaves the plate
if the nondimensional accelerational amplitude $\Gamma = a_{\rm max}/g$
exceeds unity.

After leaving the plate, the layer dilates above the plate, and then 
compresses when it collides 
with the plate later in the cycle.  When the dimensionless accelerational
amplitude $\Gamma$ is larger than a critical 
value $\Gamma_{C}$, standing-wave
patterns spontaneously form in the layer.  
These patterns are subharmonic with
respect to the plate, repeating every $2/f$ \cite{melo}.
Depending on the nondimensional accelerational amplitude $\Gamma$ and
dimensionless frequency $f^{*}=f\sqrt{H/g}$, 
various subharmonic standing waves
have been found, including stripe, square, 
and hexagonal standing-wave
patterns \cite{melo}.

\subsection{Shocks in oscillated granular layers}

If the Mach number $Ma$ (the ratio of the local mean fluid speed to
the local speed of sound) is greater than unity at the point where
a fluid encounters an obstacle, a compression wave front is formed
near the object and steepens to form a shock.  
A distinguishing feature of granular materials is that granular flows
reach supersonic speeds under common laboratory conditions 
\cite{haff, rericha, goldshtein2}.  Therefore, although shocks
are formed in ordinary gases only under extreme conditions, 
shock formation
is commonplace in granular media.

Experiments \cite{goldshtein2} and simulations 
\cite{aoki1995, potapov1996, bougie2002, carrillo2008} 
demonstrate that shock waves
form in shaken granular layers as the layer contacts the plate.
Previous investigations have generally either 
focused on shock propagation
\cite{goldshtein2, aoki1995, potapov1996} or pattern formation
\cite{melo, bougie2005, bougie2010} or considered them as 
separate phenomena coexisting in shaken layers \cite{carrillo2008}.
In this paper, we use continuum simulations
to investigate the interactions between shocks and the standing-wave 
patterns formed in this system.

\subsection{Model system}

We simulate a layer of grains on an impenetrable plate which 
oscillates sinusoidally in the direction of gravity.
The layer depth at rest is approximately $H=5.4\sigma,$ 
where the grains are modeled as identical,
frictionless spheres with diameter $\sigma$, mass $M$,
and coefficient of restitution
$e=0.7$.   In this paper, 
we study patterns and shocks as a function of 
nondimensional frequency $f^{*}$, while the 
dimensionless accelerational amplitude
$\Gamma=2.20$ is held constant.  
Previous simulations \cite{bougie2005, bougie2010} have
shown that this value exceeds the critical accelerational amplitude
$\Gamma_{C}$ indicating that we should expect to see 
standing-wave patterns for a variety of frequencies.

Previous experiments \cite{goldman03} and molecular dynamics 
(MD) simulations \cite{moon03}  
have shown that friction between grains plays a role in these patterns.
Experimentally, adding graphite to reduce
friction decreased $\Gamma_{C}$ and prevented the formation
of stable square or hexagonal patterns found for certain ranges of 
$f^{*}$ and $\Gamma$ in experiments without graphite 
\cite{goldman03}. Similarly, MD simulations with friction 
between particles have quantitatively
reproduced stripe, square, and hexagonal subharmonic standing waves seen
experimentally \cite{bizon98}, but MD simulations without 
friction yield only stable stripe patterns and
display a lower $\Gamma_{C}$ \cite{moon03}.  
In this study, we investigate stripe patterns in continuum 
simulations of frictionless particles.  To investigate other patterns
such as squares or hexagons, simulations would have to include
friction between particles.

Experiments \cite{goldman2004} and simulations 
\cite{brey2009, bougie2005, bougie2010}
indicate that fluctuations due to 
individual grain movement play a larger 
role in granular media than 
do thermal fluctuations in ordinary fluids. 
Fluctuating hydrodynamics (FHD) theory models these fluctuations 
by adding noise 
terms  to the Navier-Stokes equations 
\cite{landauandlifshitz1959, zaitsev, swifthohenberg}.  
In previous simulations with FHD noise terms, the critical value
of pattern onset $\Gamma_{C}$ was consistent with 
molecular dynamics simulations, while continuum simulations without
these fluctuations exhibited pattern onset at $\Gamma_{C}$ 
approximately $10\%$ lower than that found in MD simulations
\cite{bougie2005, bougie2010}.
Above $\Gamma_{C}$, however, simulations both with and without these
fluctuations exhibited standing-wave patterns with wavelengths
consistent with a
dispersion relation \cite{umbanhowar2000} found experimentally for 
a range of shaking frequencies.
In this paper, we do not include FHD noise terms; we investigate
patterns for accelerational amplitude greater than $\Gamma_{C}$ determined
from simulations with and without FHD.

We use continuum simulations to investigate the dynamics of 
this system including pattern formation and shock propagation.
Section~\ref{section-methods} describes the
methods we use to simulate and analyze oscillated layers.
Sec.~\ref{section-dynamics} examines the dynamics of shocks and 
standing-wave patterns formed in this system at a fixed dimensionless
oscillation frequency $f^*=0.25$,
and Sec.~\ref{section-frequency} 
examines how these patterns and shocks change when
this frequency is varied.  
We present our conclusions in Sec.~\ref{section-conclusions}.
 
\section{Methods}\rm\label{section-methods}

\subsection{Continuum equations}

We use a continuum simulation previously used to model shock waves in a
granular shaker \cite{bougie2002}.  Our simulation numerically integrates
continuum equations of Navier-Stokes order proposed by Jenkins and Richman
\cite{jenkinsandrichman} for a dense gas composed of frictionless
(smooth), inelastic hard spheres.  

This model yields hydrodynamic 
equations for number density $n$ (or equivalently, volume
fraction $\nu=\frac{\pi}{6} n\sigma^3$), velocity $\mathbf{u}$, and granular
temperature $T$:
\begin{equation} \frac{\partial n}{\partial t} +
\nabla\cdot(n\mathbf{u})=0, \end{equation}
\begin{equation}n\left( \frac{\partial\mathbf{u}}{\partial
t}+\mathbf{u}\cdot\nabla\mathbf{u} \right) =
\nabla\cdot \underline{\mathbf{P}}- n
g{\mathbf{\hat{z}}} , \label{eq:momentum}\end{equation}
\begin{equation}\frac{3}{2}n\left(\frac{\partial T}{\partial t}+
\mathbf{u}\cdot\nabla T\right) =
-\nabla\cdot \bf{q}\rm+\underline{\bf{P}}:\underline{\bf{E}}-{\gamma},
\label{eq:energy}\end{equation}
where the components of the symmetrized velocity gradient tensor
$\underline{\bf{E}}$ are given by:  $E_{ij}=\frac{1}{2}\left({\partial_j
u_i}+{\partial_i u_j}\right).$  The components of the
stress tensor $\bf\underline{P}\rm$ are given by the constitutive
relation
\begin{equation} P_{ij}=\left[ -p + (\lambda
-\frac{2}{3}\mu)E_{kk}\right] \delta_{ij}+2\mu
E_{ij},  \label{eq:stresstensor}\end{equation} 
and the heat flux is calculated from Fourier's law:
\begin{equation}\mathbf{q}=-\kappa \nabla T. \label{eq:heatflux}\end{equation}

To calculate the pressure, we use the equation of state and
radial distribution function at contact proposed by Goldshtein \it et
al. \rm \cite{goldshtein3} to
include both dense gas and inelastic effects: 
\begin{equation} p=n T \left[ 1+2(1+{e}) G(\nu)\right], \label{eq:state}\end{equation}
where
\begin{equation} G(\nu)=\nu g_0(\nu),\end{equation}
and the radial distribution function at contact, $g_0$, is
\begin{equation}g_0({\nu})=\left[ 1- \left( 
\frac{\nu}{\nu_{\rm max}}\right) ^ {\frac{4}{3}\nu_{\rm max}} \right]^{-1},\end{equation}
where $\nu_{\rm max}=0.65$ is the 3D random close-packed volume
fraction.

These equations differ from those for a compressible, dense
gas of elastic particles by the energy loss term 
\begin{equation}
\gamma = \frac{12}{ \sqrt{\pi}} (1-e^2) \frac{n T^{3/2}}{\sigma} G(\nu),
\end{equation}
which arises from the inelasticity of collisions between particles.
The bulk viscosity is given by
\begin{equation}\lambda=\frac{8}{ 3\sqrt{\pi}}n\sigma T^{1/2} G(\nu),\end{equation}
the shear viscosity by
\begin{equation}\mu=\frac{\sqrt{\pi}}{6}n\sigma
T^{1/2}\left[\frac{5}{16}\frac{1}{ G(\nu)} + 1 + 
\frac{4}{5}\left(1+\frac{12}{\pi}\right)G(\nu)\right],\label{eq:mu}\end{equation}
and the thermal conductivity by
\begin{equation}\kappa=\frac{15\sqrt{\pi}}{16}n\sigma
T^{1/2}\left[\frac{5}{24}\frac{1}{ G(\nu)} + 1 + 
\frac{6}{5}\left(1+\frac{32}{9\pi}\right)G(\nu)\right]
.\label{eq:kappa}\end{equation}

Other hydrodynamic models include modifications such as
more accurate expressions for kinetic coefficients
which incorporate high density corrections \cite{garzo1999, lutsko2005} and
equations of state which allow for volume fractions greater than 
the random-close-packed limit used here \cite{torquato1995}.  However, the 
equations shown here have previously been used to separately examine shocks 
\cite{bougie2002} and patterns \cite{bougie2005, bougie2010} in shaken layers
and have
demonstrated quantitative agreement with experiments and molecular
dynamics simulations.  Therefore, we use these equations in this study
to investigate the relationship between shocks and pattern formation; in 
principle, the equations could be modified to implement other forms of the
constitutive relations.

\subsection{Simulation method}

We integrate these
hydrodynamic equations to find number density, momentum, and granular
temperature, using a second order finite difference scheme on a uniform
grid in 3D with first order adaptive time stepping \cite{bougie2002}.
In these simulations, the granular fluid 
is contained between two
impenetrable horizontal plates at the top and bottom of the container,
where the lower plate oscillates sinusoidally between height $z=0$ and
$z=2A$ and the ceiling is located at a height $L_{z}$ above the 
lower plate.  
Periodic boundary
conditions are used in the horizontal directions $x$ and $y$ 
to eliminate sidewall effects. 
 Simulations
were conducted in a box of size $L_x=168\sigma$,
$L_y=10\sigma$, and $L_z=160\sigma$.  This orientation causes stripes 
to form parallel to the $y$-axis.
The numerical methods, boundary
conditions at the top and bottom plate, and grid spacing are the same as
used in previous studies of shocks \cite{bougie2002} 
and patterns \cite{bougie2005, bougie2010}.  

\section{Dynamics of Shocks and Standing Waves}\rm\label{section-dynamics}

\subsection{Standing wave patterns}\rm\label{subsection-patterns}

Experimental investigations of shaken granular
layers have shown that above a critical acceleration of the plate
$\Gamma_C$, standing wave patterns form spontaneously.  These patterns
oscillate subharmonically, repeating every $2/f$, so that the location of
a peak of the pattern becomes a valley after one cycle of the plate, and
vice versa \cite{melo}.  

\begin{figure}%
\center{\subfloat{\bf Volume Fraction}}\\
\subfloat{{\includegraphics[width=.4\textwidth]{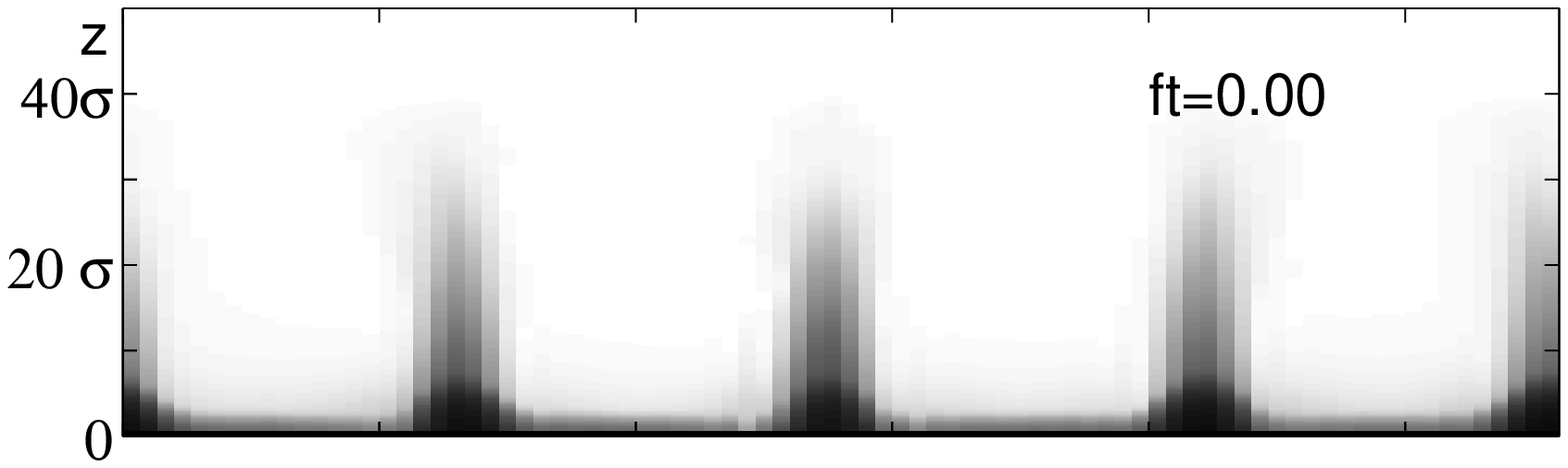}}}\\
\vspace{-0.4cm}%
\subfloat{{\includegraphics[width=.4\textwidth]{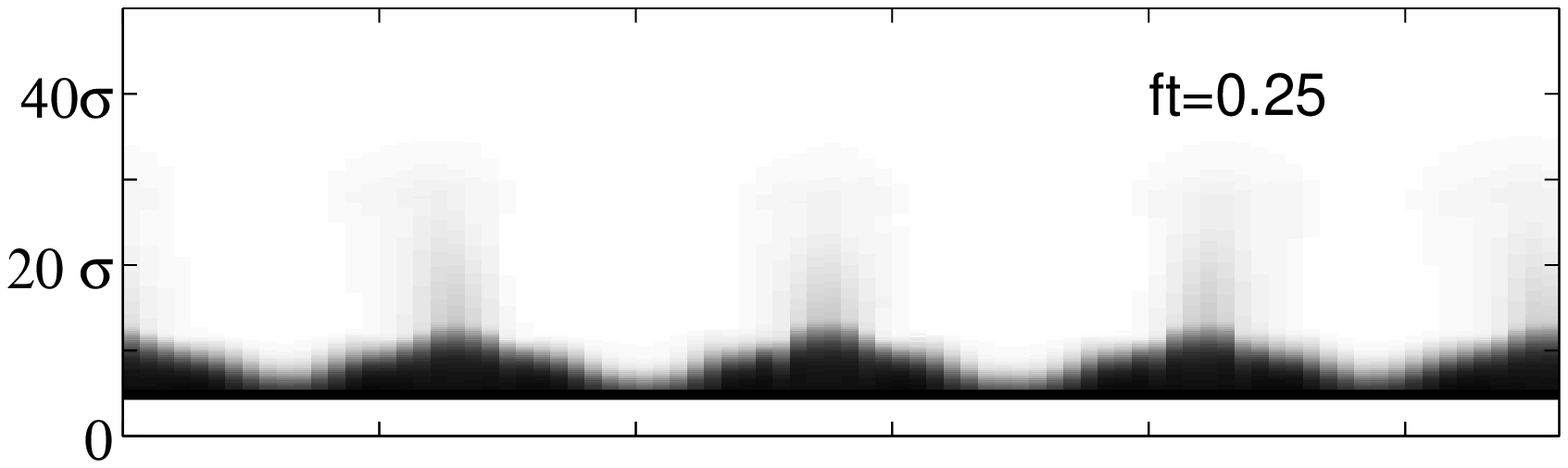}}}\\
\vspace{-0.4cm}%
\subfloat{{\includegraphics[width=.4\textwidth]{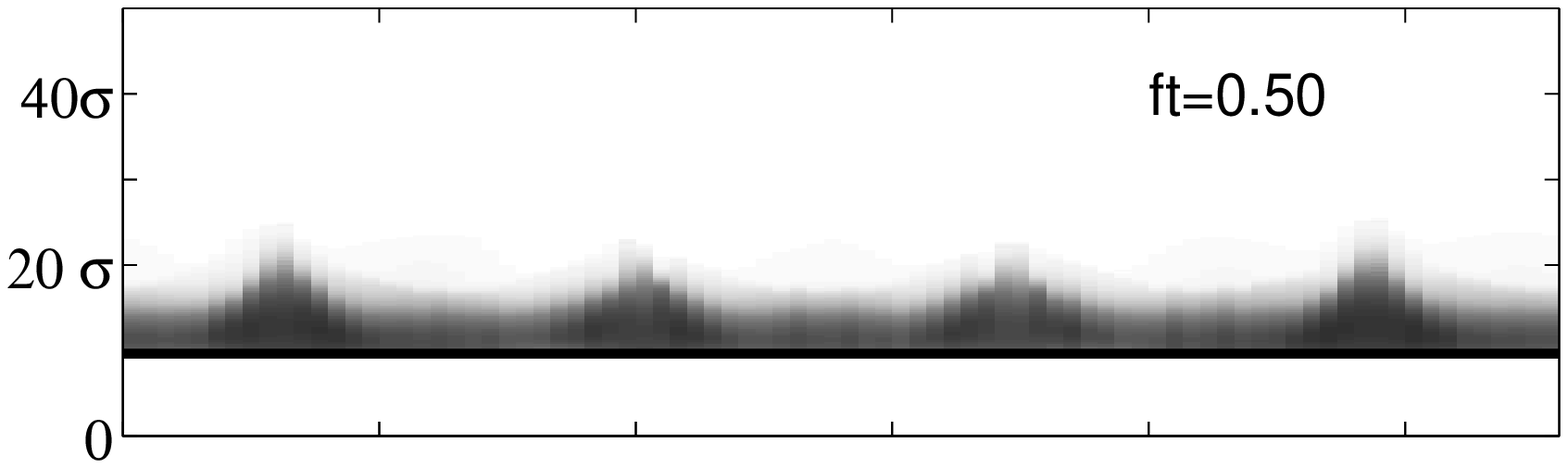}}}\\
\vspace{-0.4cm}%
\subfloat{{\includegraphics[width=.4\textwidth]{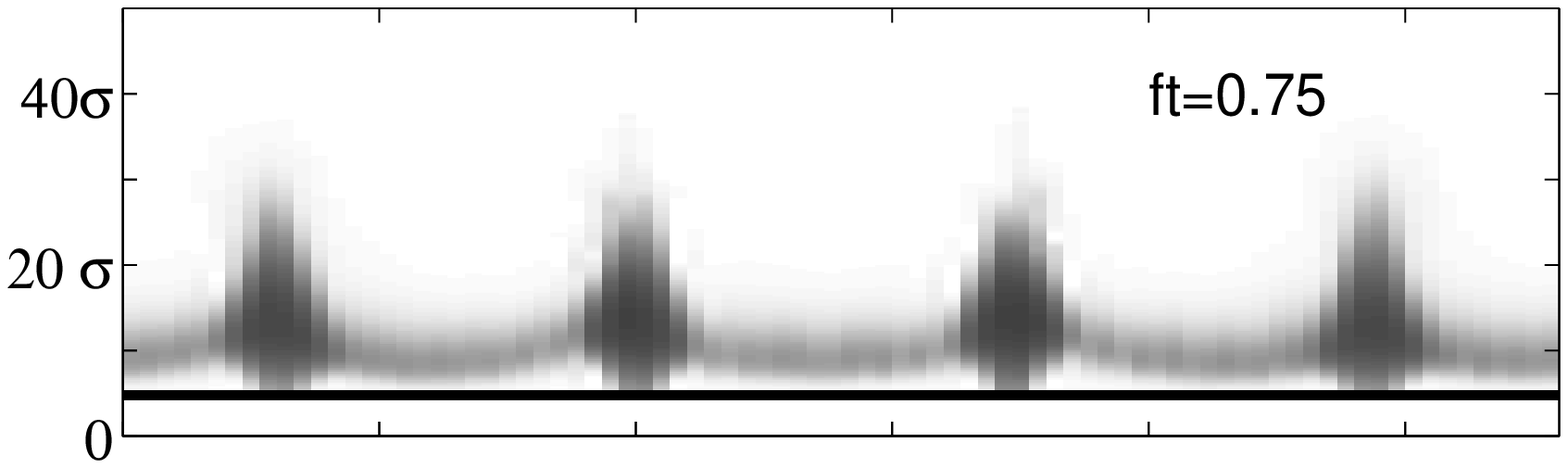}}}\\
\vspace{-0.4cm}%
\subfloat{{\includegraphics[width=.4\textwidth]{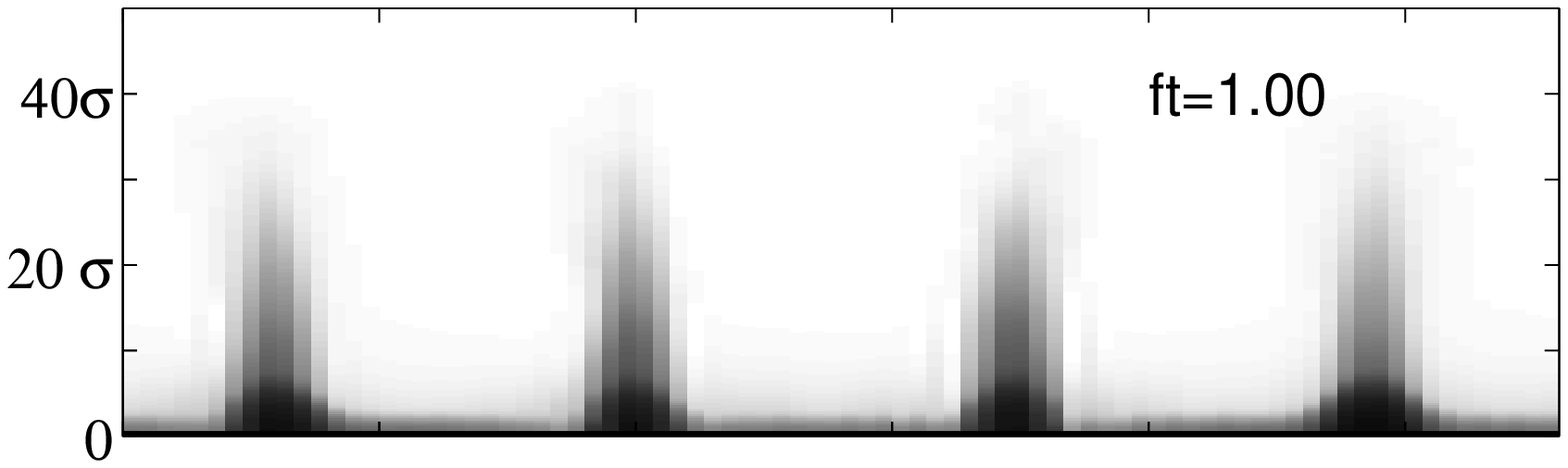}}}\\
\vspace{-0.4cm}%
\subfloat{{\includegraphics[width=.4\textwidth]{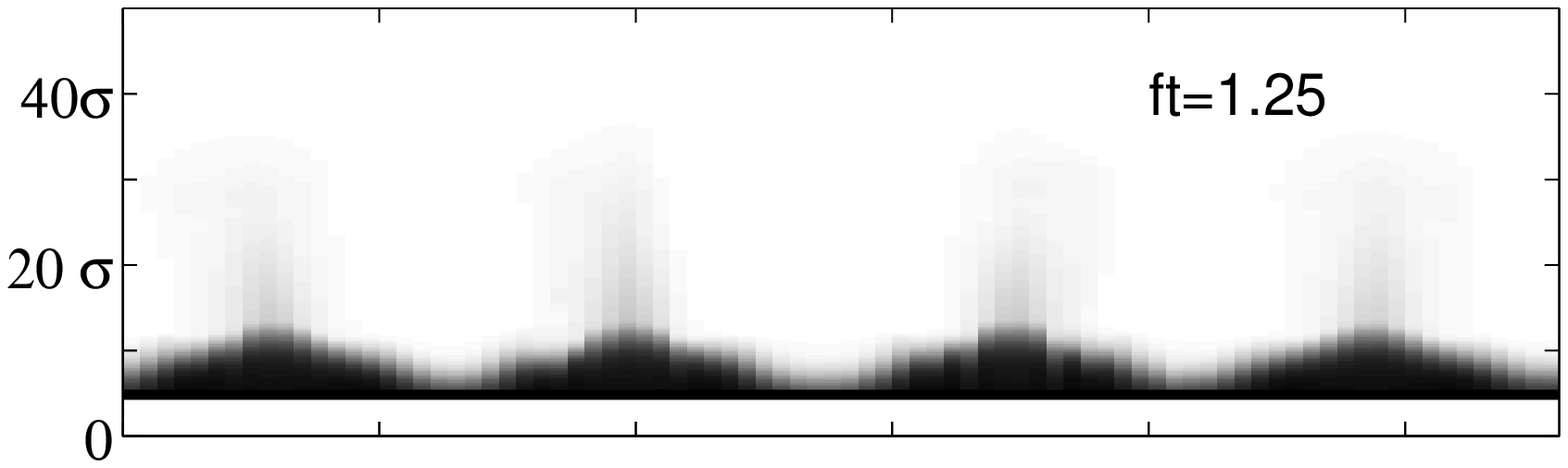}}}\\
\vspace{-0.4cm}%
\subfloat{{\includegraphics[width=.4\textwidth]{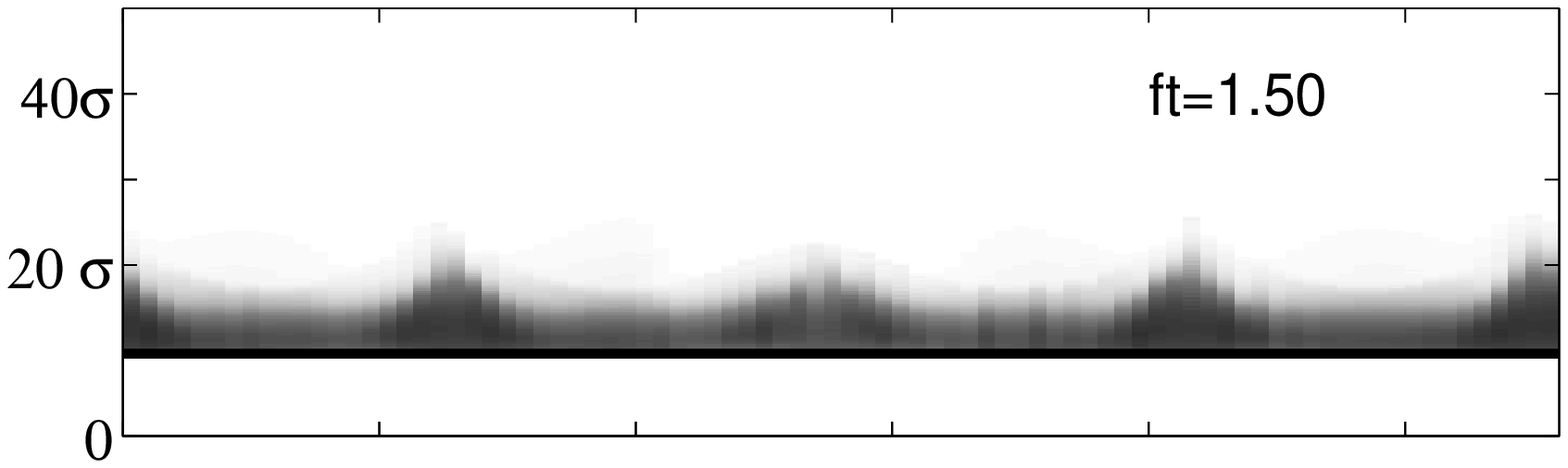}}}\\
\vspace{-0.4cm}%
\subfloat{{\includegraphics[width=.4\textwidth]{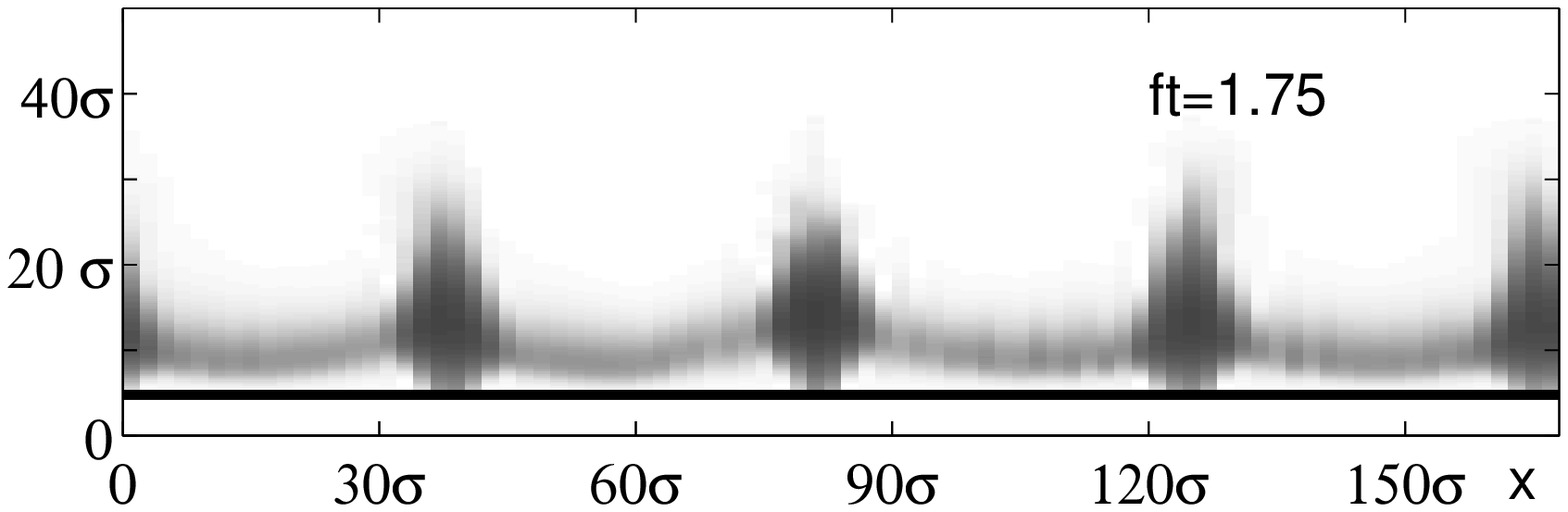}}}\\
\hspace{.5cm}{\subfloat{{\includegraphics[width=.4\textwidth]{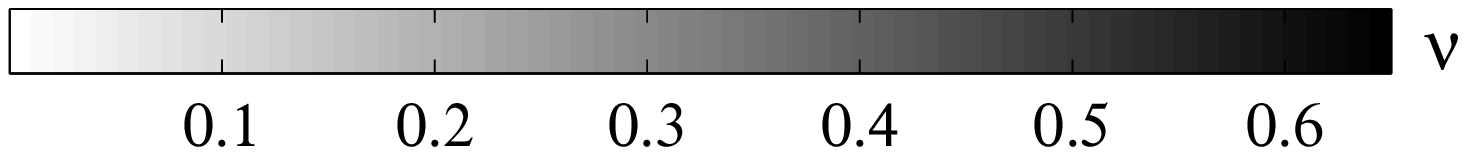}}}}\\
\vspace{-0.2cm}%

\caption{\label{patterns} 
A side view of a layer of grains, showing volume fraction $\nu$ at a slice
$y=5\sigma$ parallel to the $x - z$ plane at various times $ft$ during two
cycles of the plate.  Empty space ($\nu=0$) is white, random close-packed
volume fraction ($\nu_{\rm max}=0.65$) is black, and the color increases from white
to black through various shades of gray as volume fraction increases.
The plate is represented as a thick, horizontal black line.  The cell
extends to a height of $160 \sigma$ above the lower plate, but the figures
show only $z \lesssim 50 \sigma$, 
since the density is quite low above this height.
}
\end{figure}

Continuum simulations produce standing wave patterns for $\Gamma=2.2$ and
$f^{*}=0.25$ (Fig.~\ref{patterns}).  Beginning with a flat layer above the
plate with small amplitude random fluctuations, the simulation ran for
250 cycles of the plate until the layer reached a periodic state.  
Snapshots from various times during the
 next two cycles of the plate are shown in Fig.~\ref{patterns}.
Alternating peaks and valleys form a stripe
pattern which oscillates at $f/2$ with respect to the plate oscillation; a
location in the cell which represents a peak during one cycle will become a
valley the next cycle, and then return to a peak on the following cycle.
We examine the wave patterns at various times in the cycle $ft$ for two
cycles of the plate.

\subsubsection{$ft=0$: The layer collides with the plate.}

At $ft=0$ the container is at its minimum height.  The bulk of the layer is
dilute and four clear peaks and four valleys are evident in Fig.~\ref{patterns}
(note that one of the peaks wraps around the edges of the cell since
the boundary conditions are periodic).  The material in each peak
is falling towards the plate, 
although the bottom of the layer has begun to contact the plate and
is beginning to form a pile on the plate.  The thin layer colliding with 
the plate is characterized by a higher volume fraction $\nu$ than
the material in the peaks.

\subsubsection{$ft=0.25$: The layer piles on the plate.}

At $ft=0.25$, the plate has begun move upward as the layer falls to meet it.
At this time, most of the layer is piling on the plate, forming a high
volume fraction region near the plate.  Remnants of the peaks and valleys 
from the previous cycle are still visible, although the layer is
flattening on the plate, with the height difference between peaks and
valleys much smaller than at $ft=0$ (Fig.~\ref{patterns}).

\subsubsection{$ft=0.50$: The layer begins to leave the plate.}

As the container reaches its maximum downward acceleration at the top of its
oscillation, the layer begins to leave the plate.  Although the bottom of the
layer is still in contact with the plate at this time, the layer is expanding,
as evidenced by the fact that it is 
visibly more dilute in this picture 
than in the previous snapshot (Fig.~\ref{patterns}).  
As the layer leaves the plate, 
new peaks and valleys develop, 
with peaks forming in the valleys from the
previous cycle, and vice versa.

\subsubsection{$ft=0.75$: The layer is off the plate.}

At $ft=0.75$, the plate is moving downward leaving a gap between the
layer and the plate.  The mass of the layer has almost entirely left 
the plate, and the layer has expanded.  The new peaks and valleys have
become quite distinct, with a large portion of the layer in the 
peaks and very little material in the valleys connecting
them (Fig.~\ref{patterns}).

\subsubsection{$1.0\lesssim ft \lesssim 2.0$: The cycle repeats.}

At $ft=1.0$, the plate has undergone one full oscillation and has
returned to its lowest point in the cycle.  The features of this point in
the cycle are similar to those at $ft=0$, except the peaks and valleys
have reversed location.  Aside from the peaks and valleys reversing, the
next cycle exhibits the same features as the previous cycle at all 
times during the cycle.  By $ft=1.75$ the peaks and valleys are
back to their original location, demonstrating the subharmonic nature
of these patterns.

\subsection{Shocks}\label{subsection-shocks}

Previous experiments  
\cite{goldshtein2} and simulations
\cite{aoki1995, potapov1996, bougie2002, carrillo2008} 
indicate that shocks are formed in vertically oscillated granular
layers during each collision of the layer with the plate.
A prerequisite for
shock formation is that the local Mach number of the flow be greater than
unity with respect to the object causing the disturbance.  Therefore,
we calculate the speed of sound using a relation derived from
the equation of state Eq.~(\ref{eq:state}) \cite{savage}:
\begin{equation}c=\sqrt{T\chi\left(
1+\frac{2}{3}\chi+\frac{\nu}{chi}\frac{\partial\chi}{\partial\nu}\right)
},\end{equation}
where $\chi = 1+2(1+e)G(\nu)$.  The Mach number $Ma$ is calculated
as the local flow speed divided by the local speed of sound 
$Ma=\left|{\bf u }\right|/c$.

\begin{figure*}[htb]%
\center{
\subfloat{\bf Horizontal Location $x=16\sigma$\nonumber}
\hspace{4 cm}\subfloat{\bf Horizontal Location $x=32 \sigma$}}\\
\subfloat{\label{oneda}\scalebox{0.55}{{\includegraphics{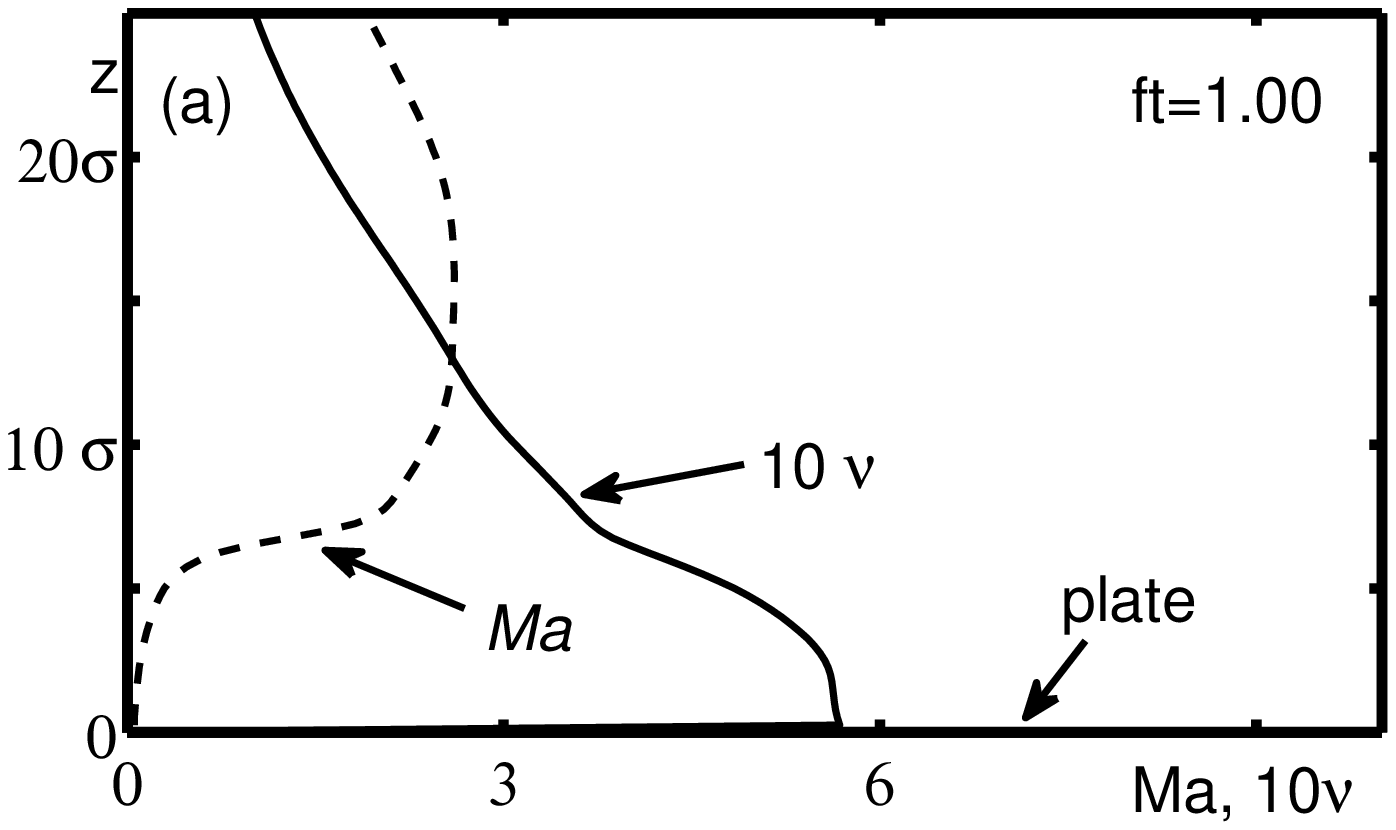}}}}
\hspace{-0.7cm}%
\subfloat{\label{onedb}\scalebox{0.55}{{\includegraphics{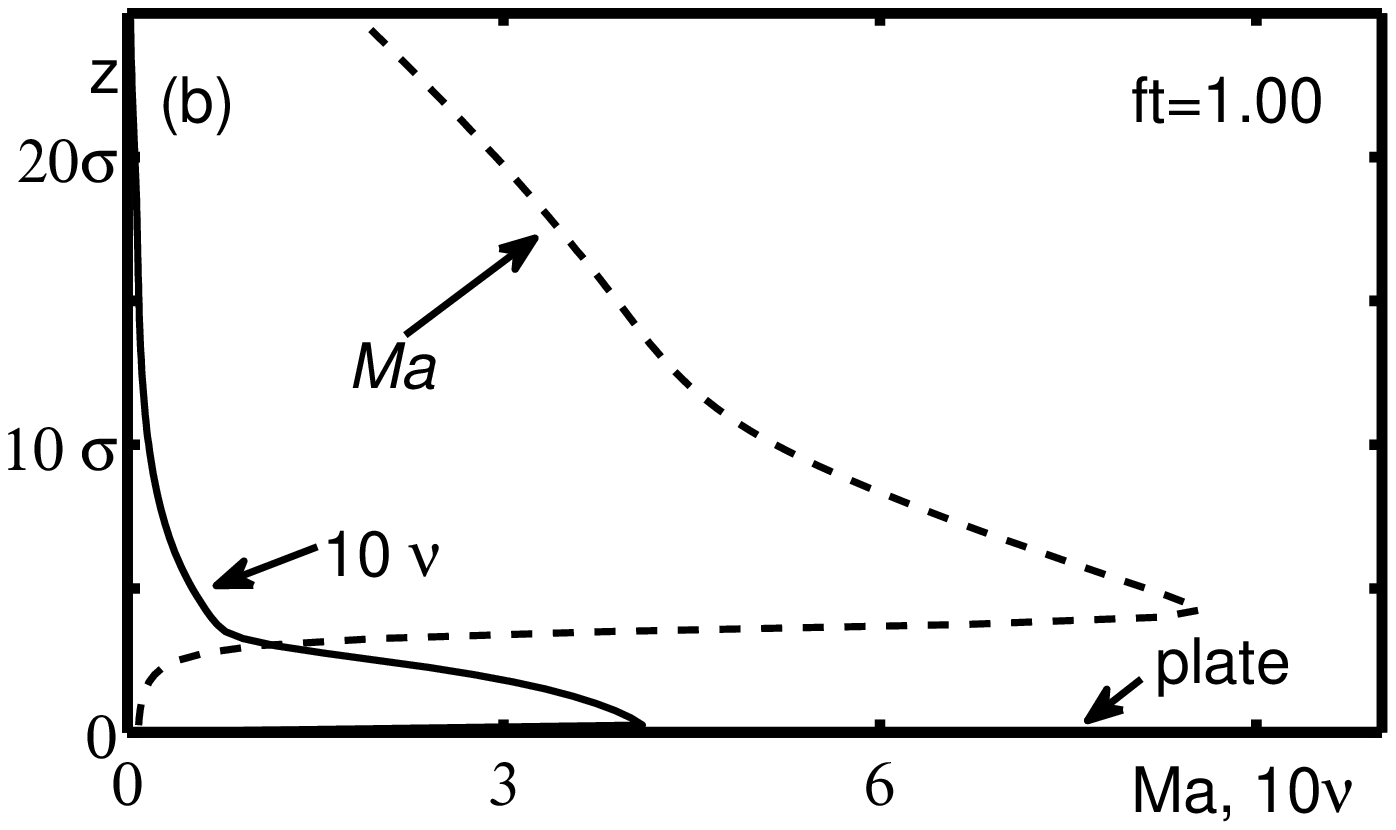}}}}\\
\vspace{-0.8cm}%
\subfloat{\label{onedc}\scalebox{0.55}{{\includegraphics{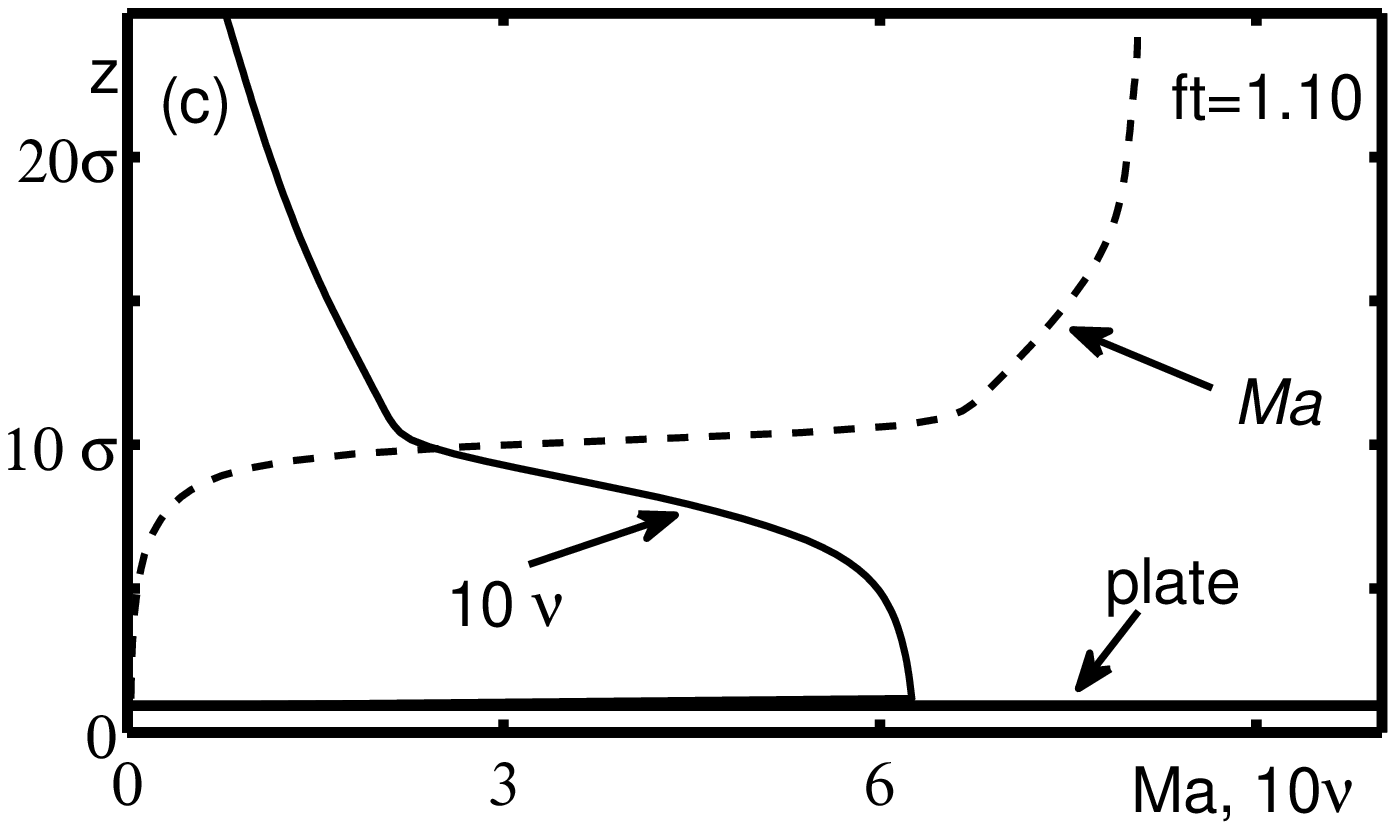}}}}
\hspace{-0.7cm}%
\subfloat{\label{onedd}\scalebox{0.55}{{\includegraphics{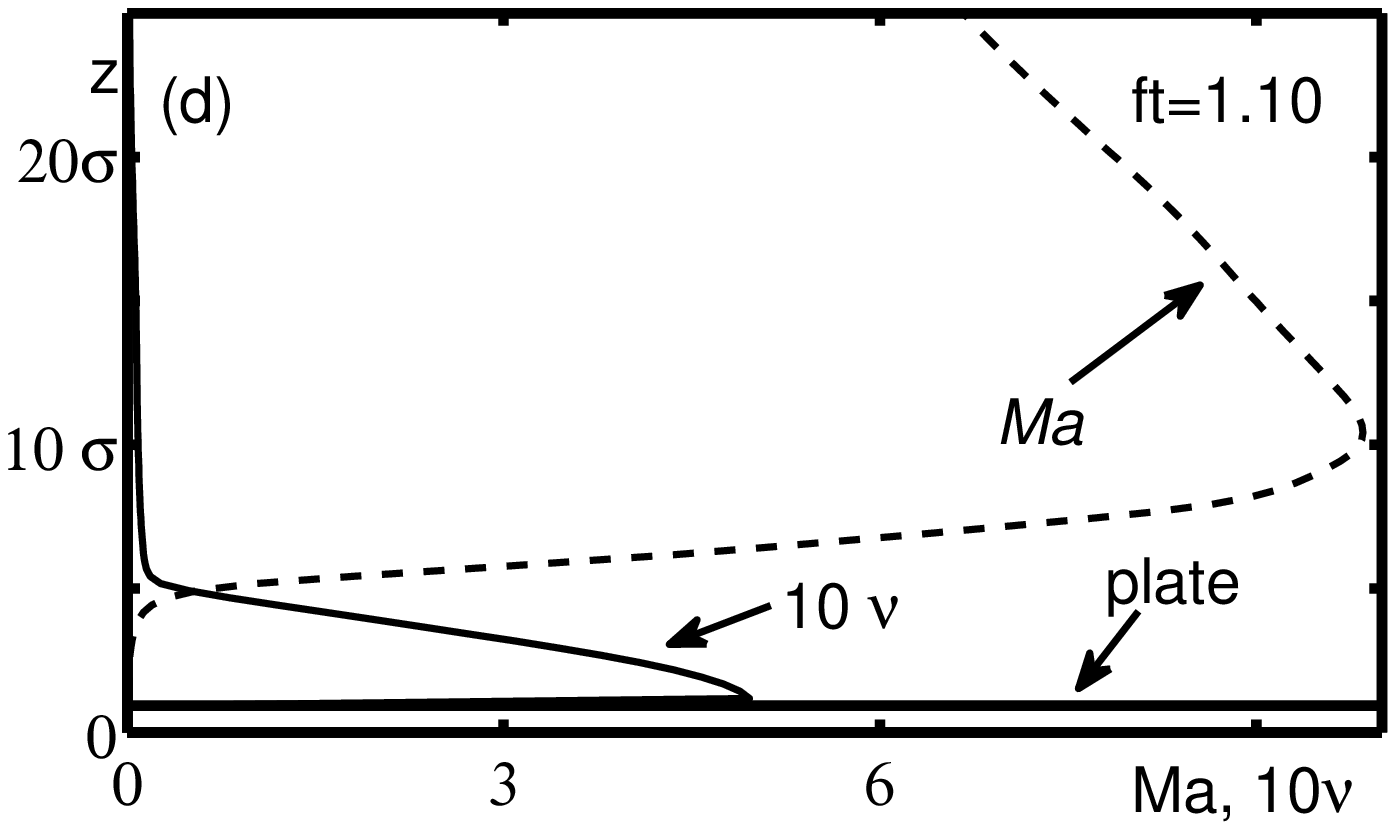}}}}\\
\caption{\label{onedshocks} 
Mach number $Ma$ (dashed line) and rescaled volume fraction $10 \nu$ 
(solid line)
as functions of height $z$ (ordinate) at two times $ft$ in the oscillation
cycle.  In each case, the plate is shown as a horizontal solid black line.
The left column shows $Ma$ and $10 \nu$ at the horizontal location 
$y=5 \sigma$, $x=16\sigma$ at times (a) $ft=1.0$ and (c) $ft=1.1$ in
the plate's oscillation, while the right column shows
 $Ma$ and $10 \nu$ at the horizontal location 
$y=5 \sigma$, $x=32\sigma$ at times (b) $ft=1.0$ and (d) $ft=1.1$.
A shock is present in each case, as indicated by a discontinuity in 
the volume fraction derivative and a sharp increase in $Ma$ moving
away from the plate.}
\end{figure*}

From Fig.~\ref{patterns}, we see that at $ft=1.00$, 
a portion of the layer
has begun to compress on the plate, while some of the material 
is still falling towards the plate.  This is therefore
a time at which we would expect to see shock formation.
To examine the shock in detail, we plot one-dimensional
profiles of volume fraction $\nu$ and Mach number $Ma$ 
(here $\nu$ is scaled by a factor of 10 to fit on the 
same scale as $Ma$) as 
functions of height $z$ from the plate in Fig.~\ref{onedshocks}.

At $ft=1.00$, the first peak of the pattern occurs at the horizontal
location $16\sigma\lesssim x \lesssim 20\sigma$ 
({\it cf} Fig.~\ref{patterns}).  
We choose two horizontal locations
to examine; the location $y=5\sigma$, and $x=16\sigma$
is shown in the left column of Fig.~\ref{onedshocks}, corresponding
to the first peak of the pattern; the location $y=5\sigma$, 
and $x=32 \sigma$ is shown in the right column, corresponding to
a location away from the peak.

At the horizontal location
$y=5 \sigma$, $x=16\sigma$ corresponding to a peak of the
wave pattern, the volume fraction
approaches the close-packed value $\nu \approx 0.6$ 
near the plate  at $ft=1.00$ (Fig.~\ref{onedshocks}a).
As height increases, $\nu$ smoothly decreases for
$0\leq z \lesssim 7\sigma$.  The Mach 
number with respect to the plate 
$Ma$ is low throughout this
region, as this part of the layer is compressed on the plate and
thus matches the velocity of the plate at height $z=0$. 
However, at $z \approx 7 \sigma$, there
 is a discontinuity in the derivative of $\nu$, and a sharp increase
in Mach number such that  $Ma\approx3$ for $z\gtrsim 8\sigma$.  This
sharp increase in Mach number from subsonic to supersonic, corresponding
to a discontinuity in the derivative of volume fraction at the same
height, indicates that this is the location of a strong shock front.

While the above description applies to $x=16\sigma$ (the horizontal
location of the first peak), a shock also forms in the valleys.
This can be seen in Fig.~\ref{onedshocks}b, which shows $10 \nu$
and $Ma$ at the same time, but at horizontal location $x=32\sigma$.
Here we again see a high density, subsonic region near the plate,
with a shock separating this region from a lower density, supersonic
undisturbed region that is still falling towards the plate.
However, the volume fraction near the plate
is smaller than it was in  Fig.~\ref{onedshocks}a, 
and the shock profile is quite different.  Also, the 
shock front at $x=32\sigma$ is closer to the plate than 
it is at $x=16\sigma$.

By $ft=1.10$, the shock has moved away from 
the plate and it continues to 
develop as it moves through different parts of the layer 
(Fig.~\ref{onedshocks}c).
The shock front in the valleys is also developing 
and moving away from the 
plate (see Fig.~\ref{onedshocks}d), although
the location and profile of the shock is different at these different
horizontal locations.  

Thus, shocks form with each collision of the layer with the
plate, and the shock profile changes and develops as the
shock moves through the layer.  Shocks form at
each horizontal location in the cell; however 
the shock front is non-uniform horizontally due to the 
differences in layer depth between the peaks and the valleys
of the pattern.

\subsection{Interaction between shocks and patterns}

There is therefore a relationship between the shocks
and the standing-wave patterns formed in the layer. 
If the layer were perfectly flat,
the shock front would be uniform horizontally.  However, as there is 
horizontal variation in layer depth, a non-uniform shock front forms.

While the system is driven by vertical motion of the plate,
the patterns are characterized by horizontal variation between
peaks and valleys.  Periodic horizontal 
motion is required to produce the subharmonic oscillation.  
This horizontal
flow can be thought of as a ``sloshing'' motion, 
in which material from 
the layer flows out of some regions and into others, 
creating peaks and valleys.
With the next collision of the layer
with the plate, the flow direction reverses, 
creating peaks where there were valleys, and
vice versa.
In this section, 
we show that strong pressure
gradients created by the shock drive the sloshing
motion of the layer.

\begin{figure*}[htb]%
\center{
\hspace{.5 cm}\subfloat{\bf Volume Fraction}
\hspace{3.5 cm}\subfloat{\bf Dimensionless Pressure}}\\
\subfloat{\scalebox{0.45}{{\includegraphics{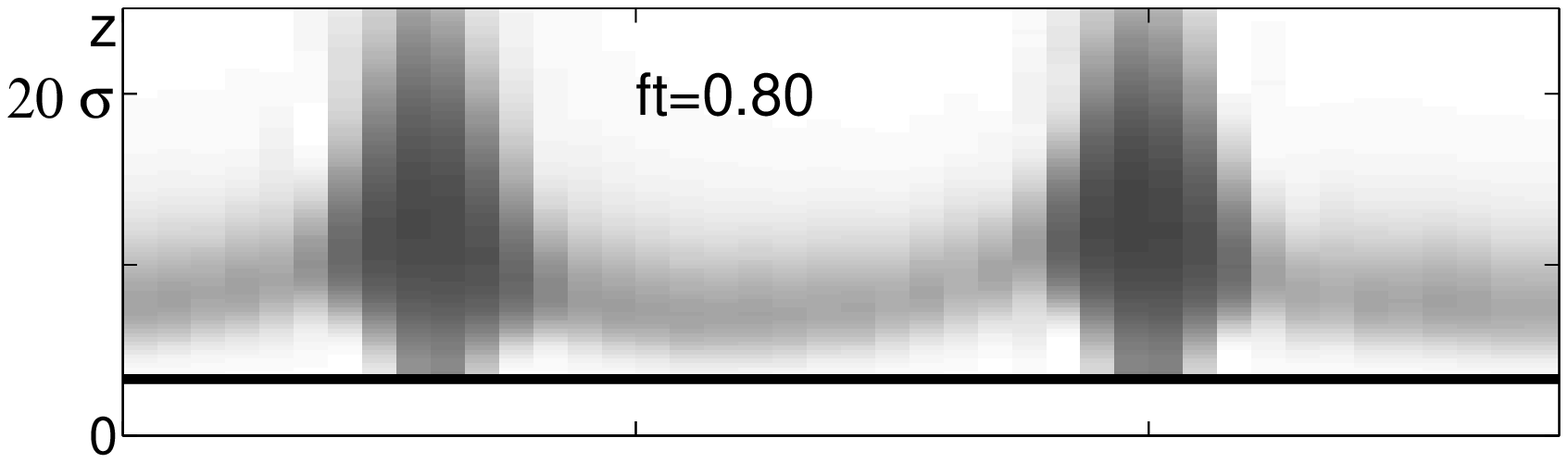}}}}
\subfloat{\scalebox{0.45}{{\includegraphics{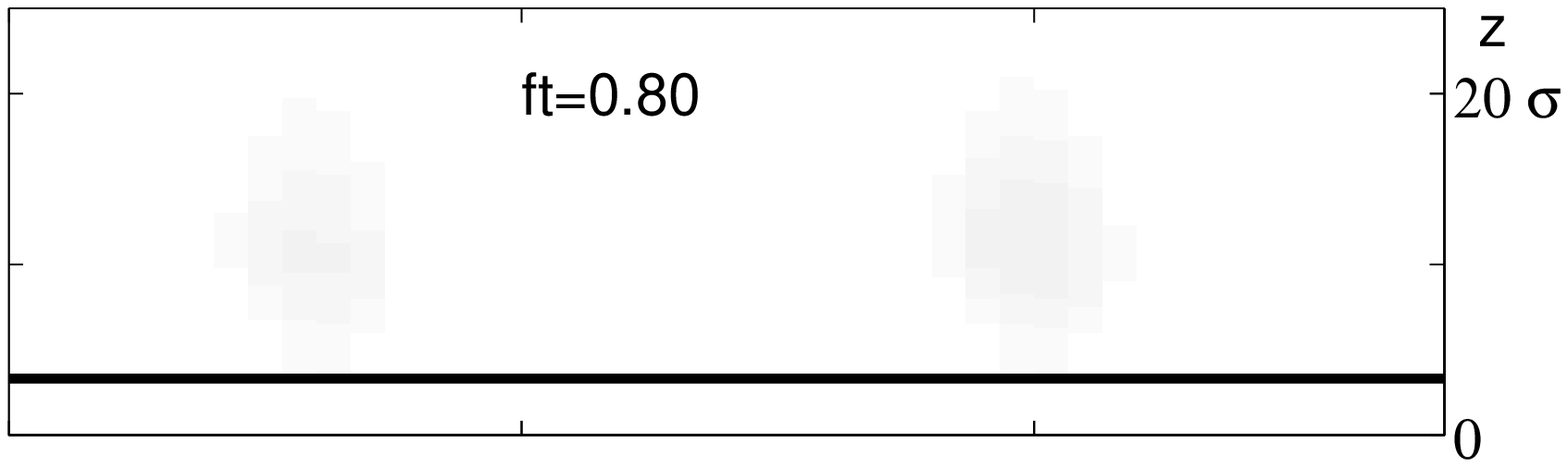}}}}\\
\vspace{-0.45cm}%
\subfloat{\scalebox{0.45}{{\includegraphics{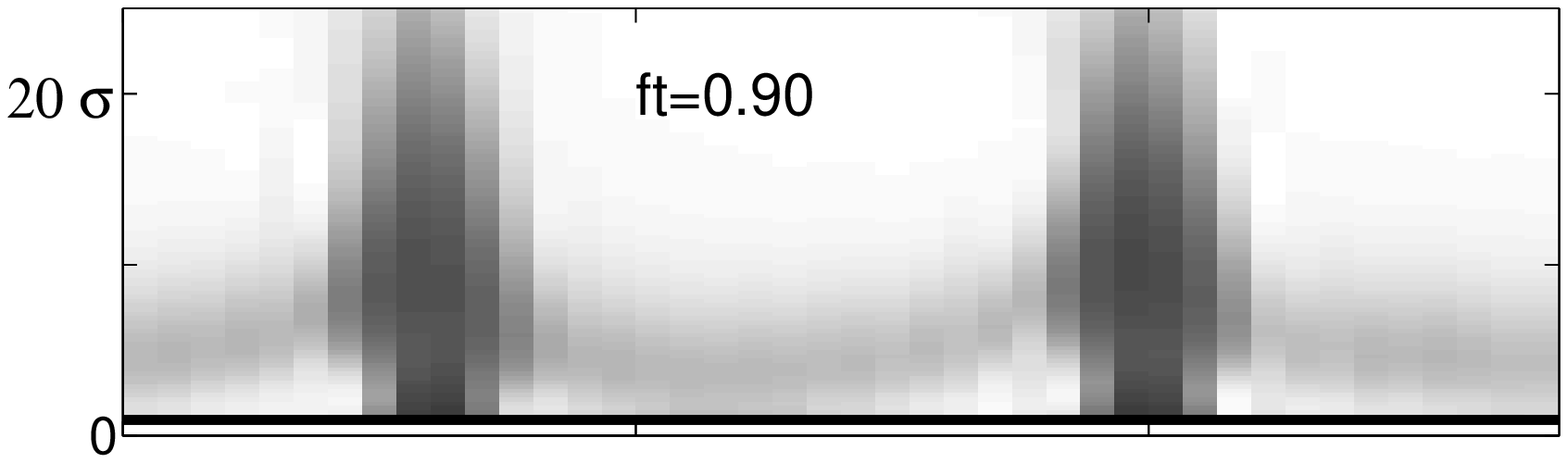}}}}
\subfloat{\scalebox{0.45}{{\includegraphics{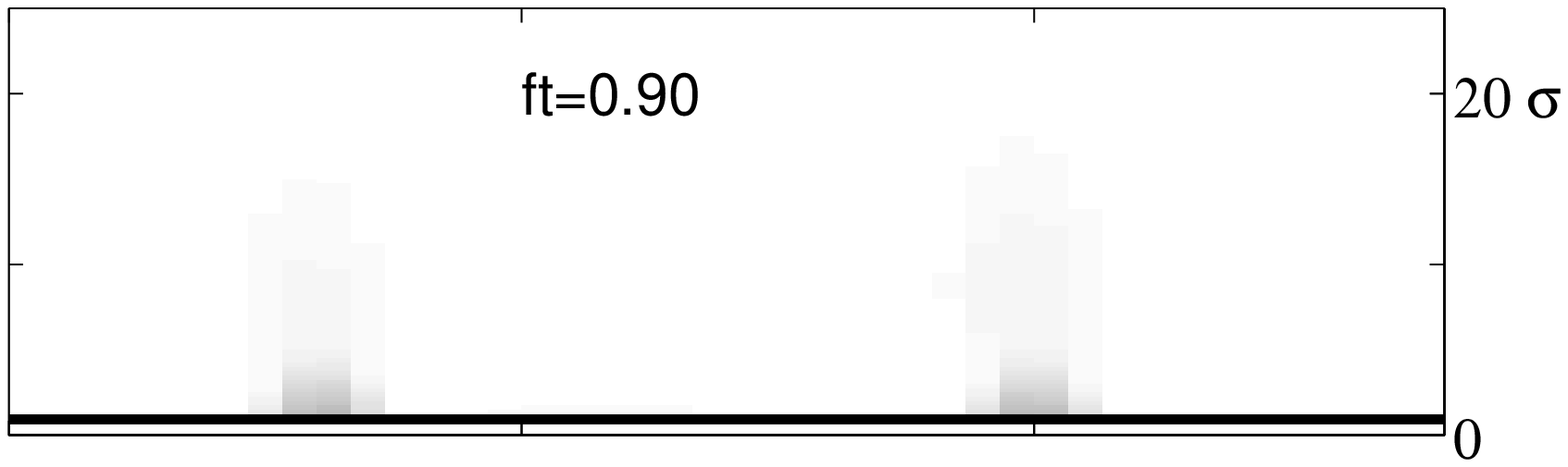}}}}\\
\vspace{-0.45cm}%
\subfloat{\scalebox{0.45}{{\includegraphics{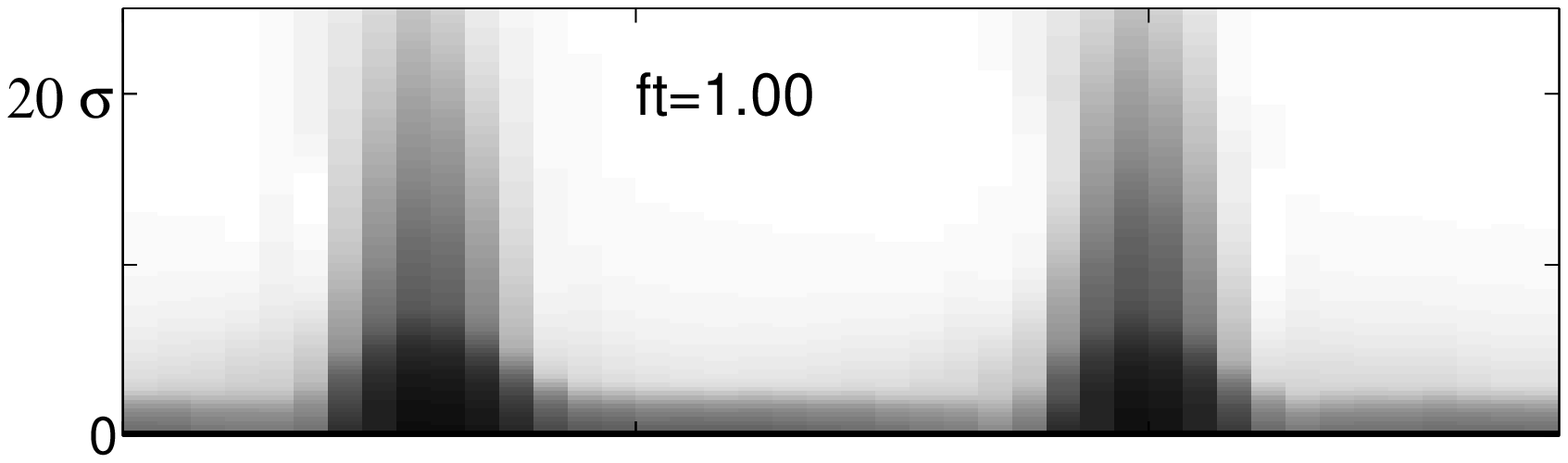}}}}
\subfloat{\scalebox{0.45}{{\includegraphics{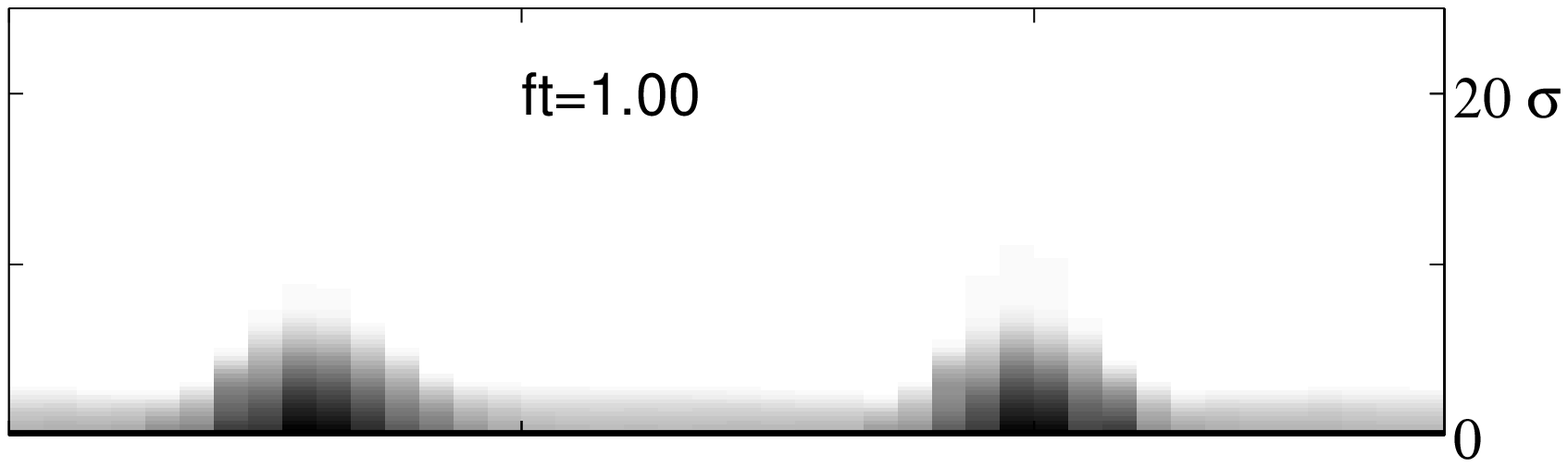}}}}\\
\vspace{-0.45cm}%
\subfloat{\scalebox{0.45}{{\includegraphics{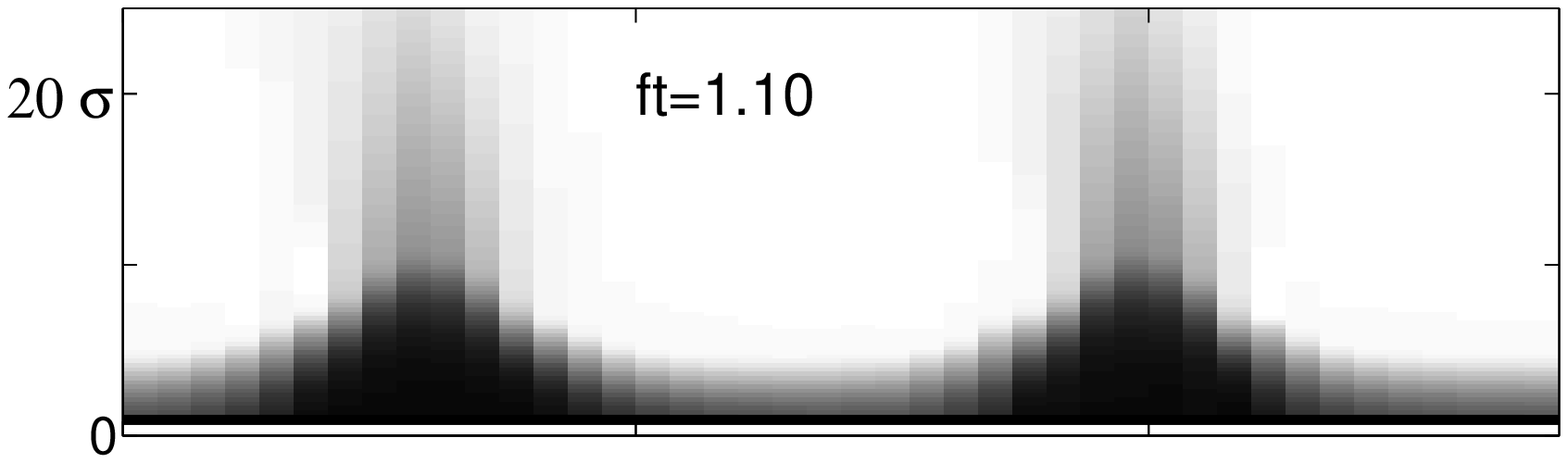}}}}
\subfloat{\scalebox{0.45}{{\includegraphics{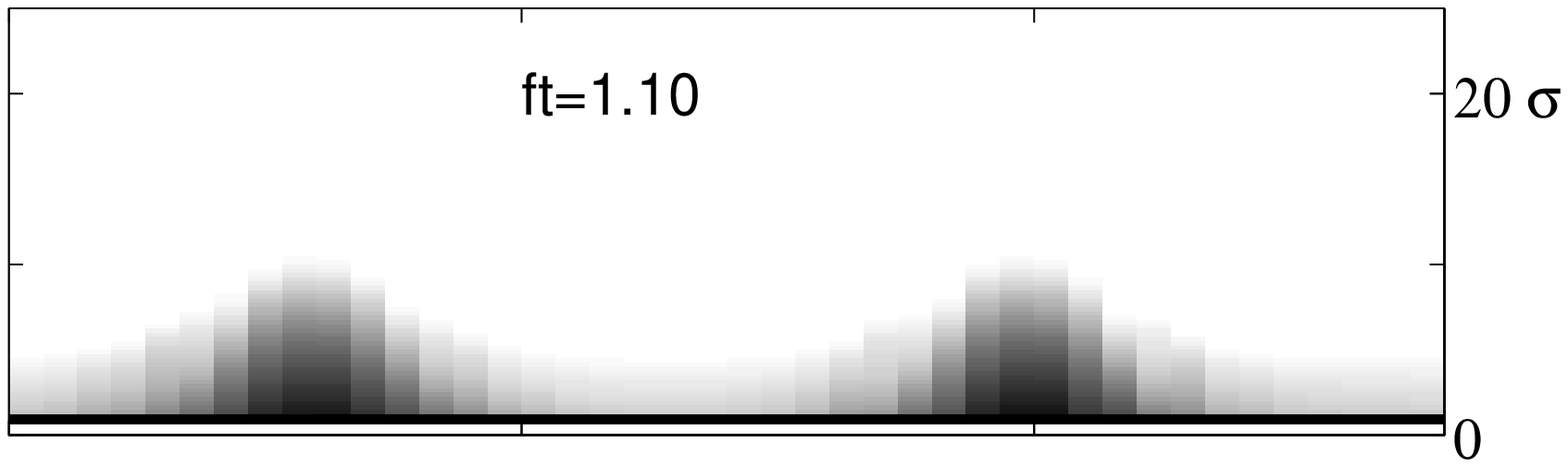}}}}\\
\vspace{-0.45cm}%
\subfloat{\scalebox{0.45}{{\includegraphics{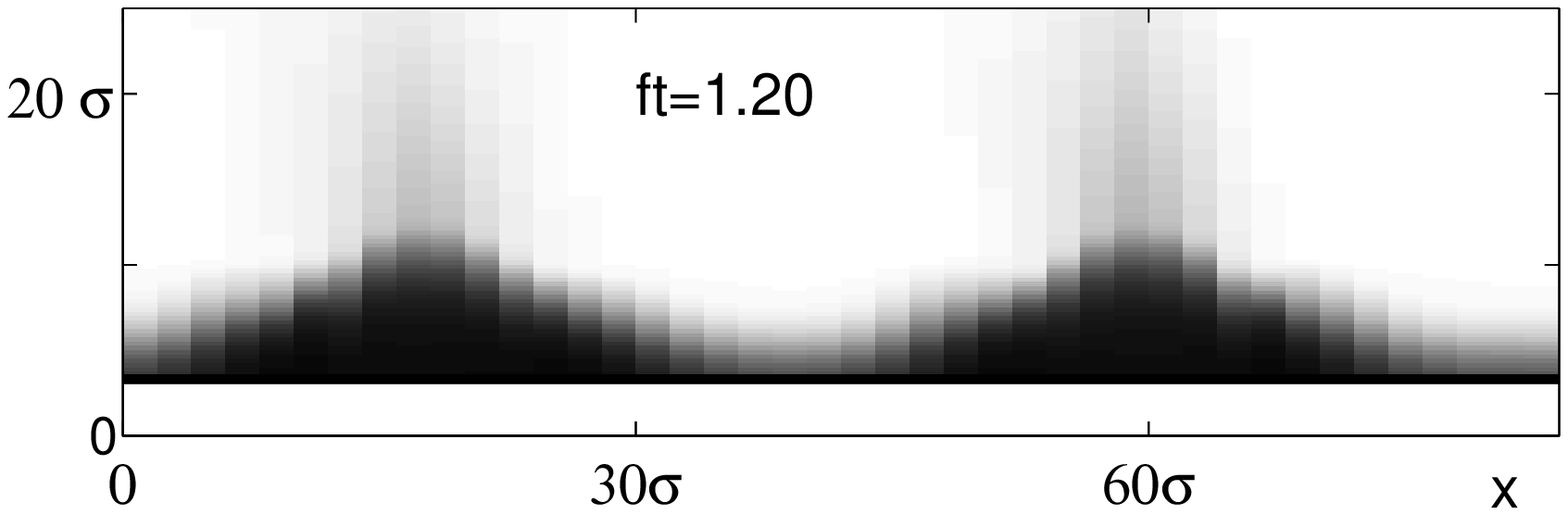}}}}
\subfloat{\scalebox{0.45}{{\includegraphics{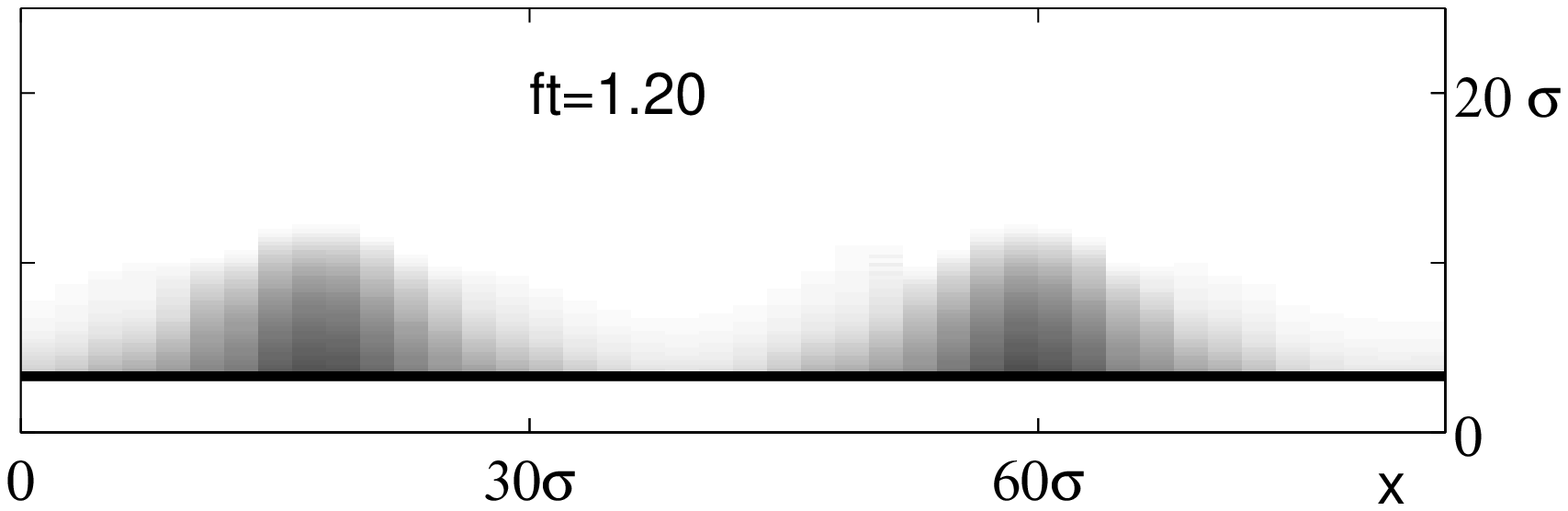}}}}\\
\hspace{1.2cm}
\subfloat{\scalebox{0.45}{{\includegraphics{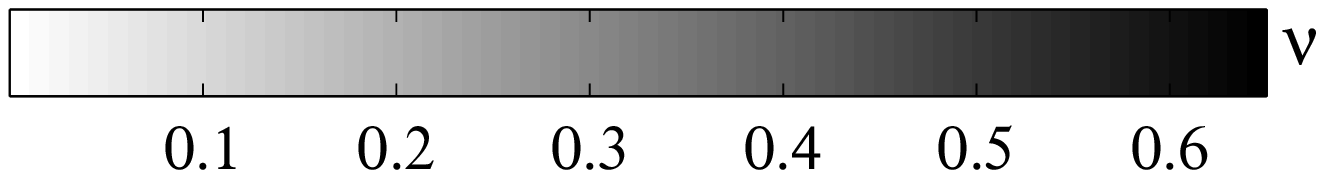}}}}
\hspace{-0.6cm}
\subfloat{\scalebox{0.45}{{\includegraphics{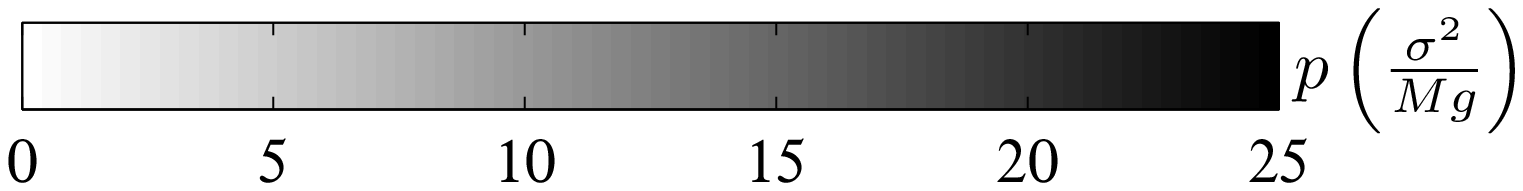}}}}\\
\caption{\label{shocks} A side view of a layer of grains, 
showing volume fraction $\nu$ (left column) and dimensionless pressure
$p\left(\frac{\sigma^2}{M g}\right)$ (right column), at a slice
$y=5\sigma$ parallel to the $x - z$ plane at various times $ft$.  
Although the full cell has a length
$168\sigma$ in the $x$-direction and a height $160\sigma$ in the 
$z$-direction, this figure only displays the horizontal range
$0\leq x\leq84\sigma$ and the vertical range $0\leq z\leq 25\sigma$
to show a closer view of the collision of the shock with the plate.
In the left column, empty space ($\nu=0$) is white, random close-packed
volume fraction ($\nu=0.65$) is black, and the color increases from white
to black through shades of gray as volume fraction increases.
The plate is represented as a thick, horizontal black line.  
In the right column, low dimensionless pressure is white, 
high dimensionless pressure is black, and pressure increases from 
white to black through shades of gray.}
\end{figure*}

To study this interaction, we investigate the layer properties 
near the time of shock formation.
According to Fig.~\ref{patterns}, 
most of the layer is off the plate at $ft=0.75$,
while the layer is compressed on the plate and the peaks and
valleys have flattened significantly by $ft=1.25$.  
In Fig.~\ref{shocks} we show 
snapshots of the the volume
fraction $\nu$ (left column) and the dimensionless pressure 
$p\left(\frac{\sigma^2}{M g}\right)$
(right column) at five different times 
$0.8\leq ft\leq 1.2$.  
We display side views of the layer at $y=5\sigma$
in  (Fig~\ref{shocks}), showing
only the horizontal 
range $0 \leq x \leq 84\sigma$
and the vertical range $0 \leq z \leq 25\sigma$
to better view the interaction
between the layer and the plate.  
This range shows
two wavelengths of the standing wave pattern and 
corresponds to the bottom
left quadrant of the range plotted in Fig.~\ref{patterns}.  

\begin{figure*}[htb]%
\center{
\hspace{-1.5 cm}
\subfloat{\bf Dimensionless Horizontal Pressure Gradient}}
\hspace{.75 cm}
\subfloat{\bf Dimensionless Horizontal Velocity}\\

\subfloat{\scalebox{0.45}{{\includegraphics{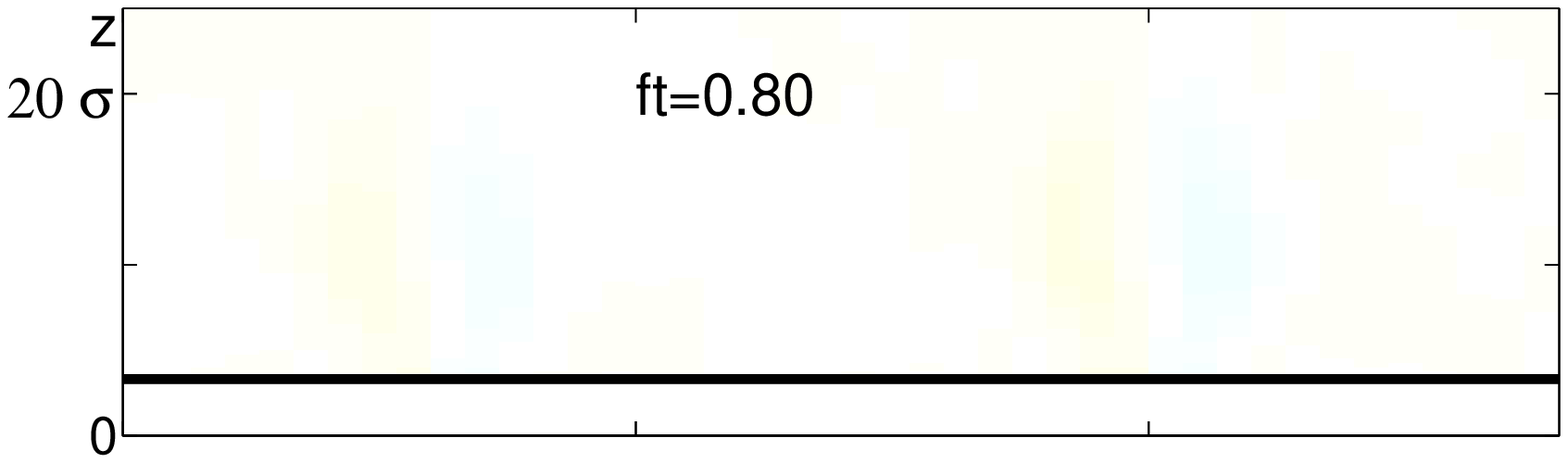}}}}
\subfloat{\scalebox{0.45}{{\includegraphics{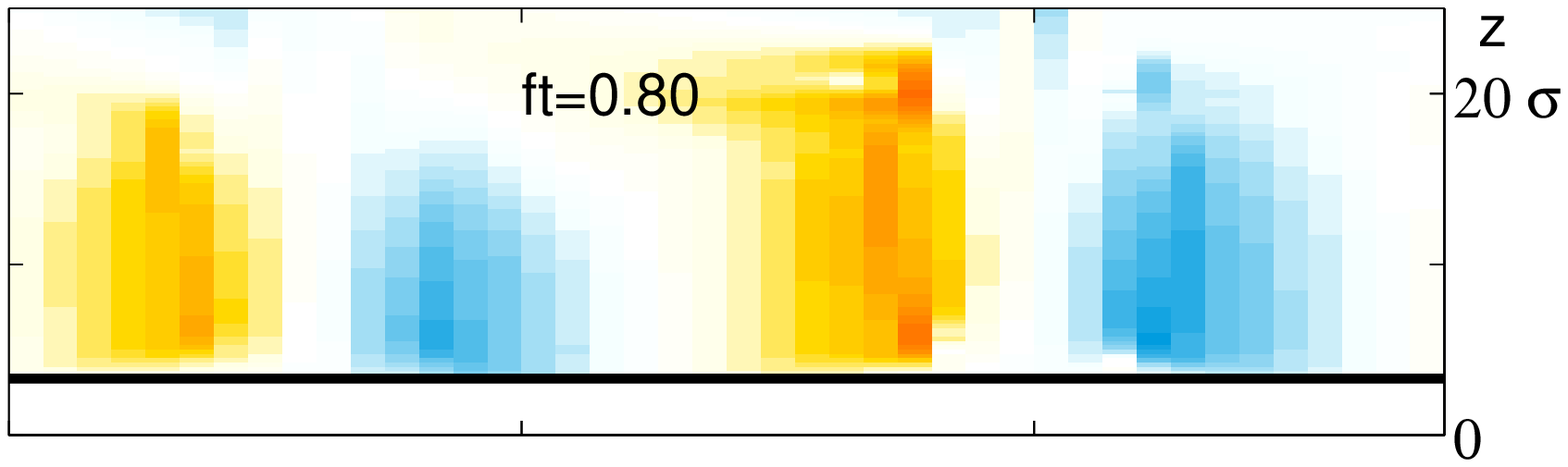}}}}\\
\vspace{-0.45cm}%
\subfloat{\scalebox{0.45}{{\includegraphics{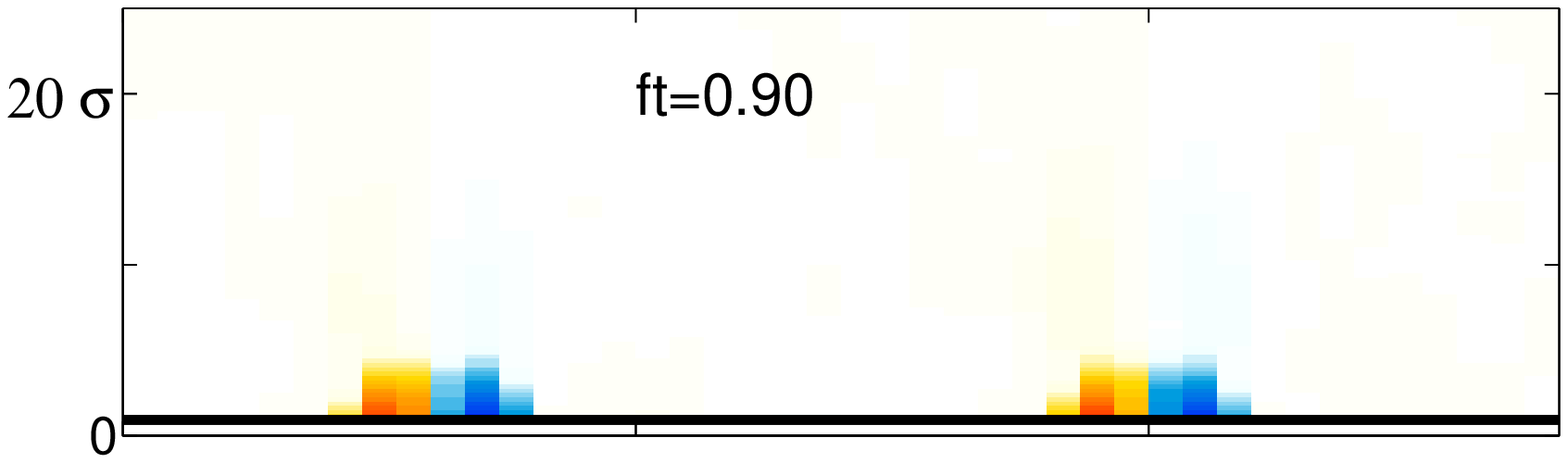}}}}
\subfloat{\scalebox{0.45}{{\includegraphics{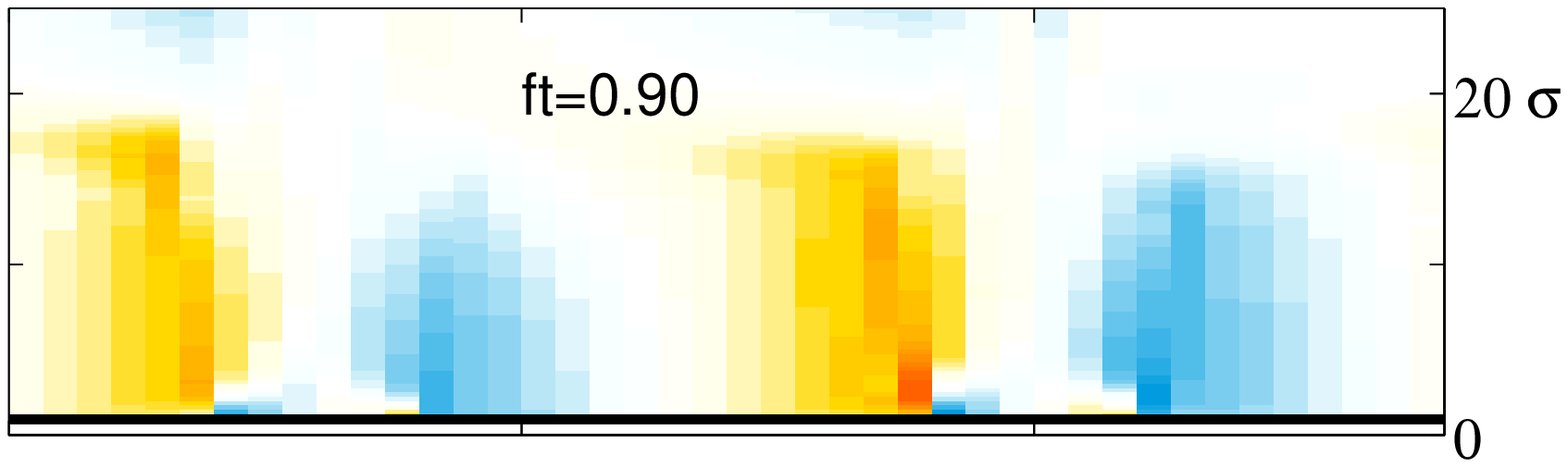}}}}\\
\vspace{-0.45cm}%
\subfloat{\scalebox{0.45}{{\includegraphics{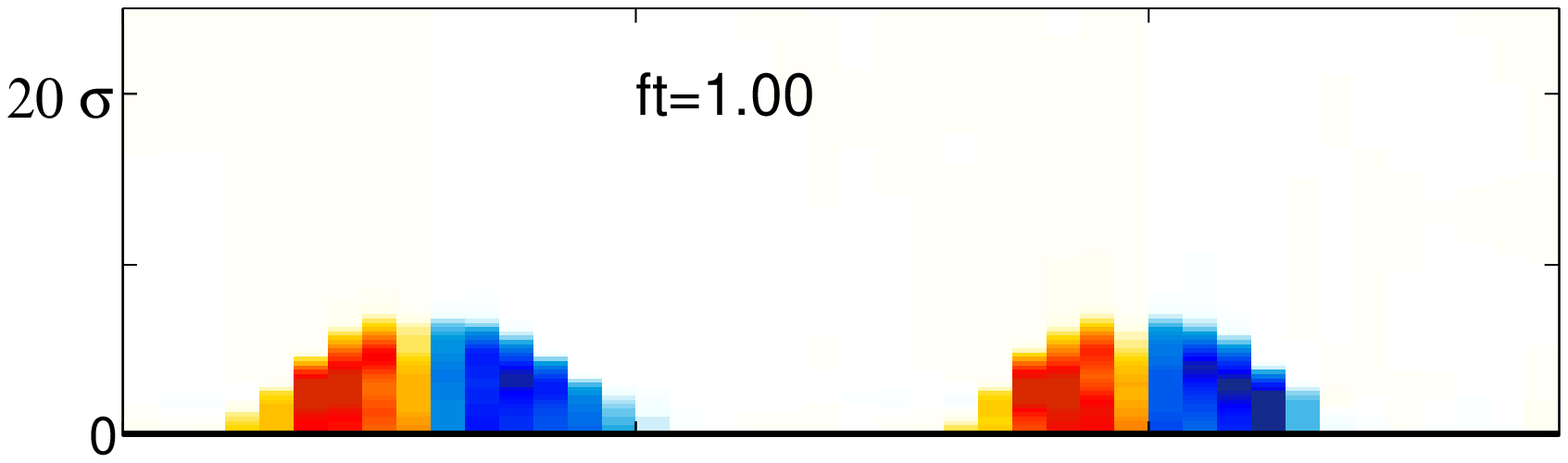}}}}
\subfloat{\scalebox{0.45}{{\includegraphics{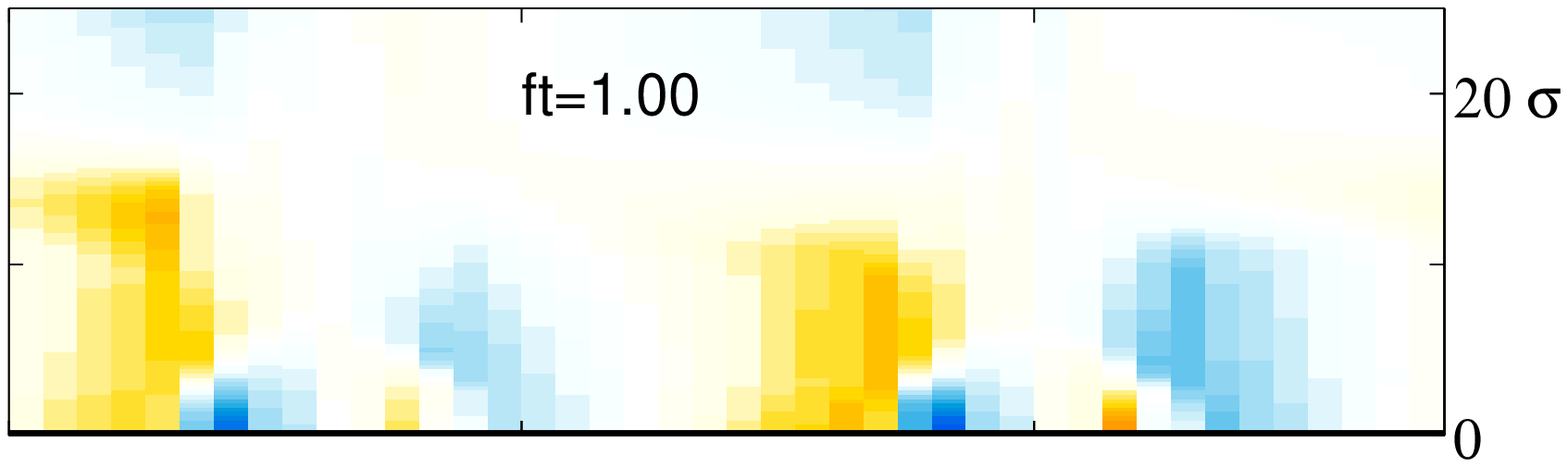}}}}\\
\vspace{-0.45cm}%
\subfloat{\scalebox{0.45}{{\includegraphics{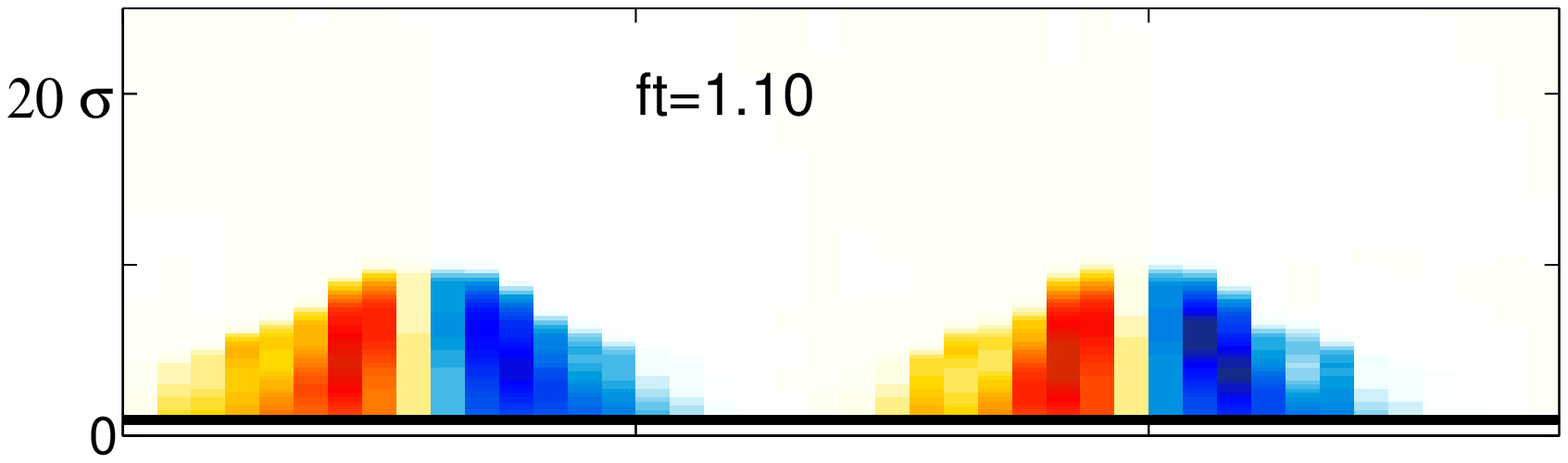}}}}
\subfloat{\scalebox{0.45}{{\includegraphics{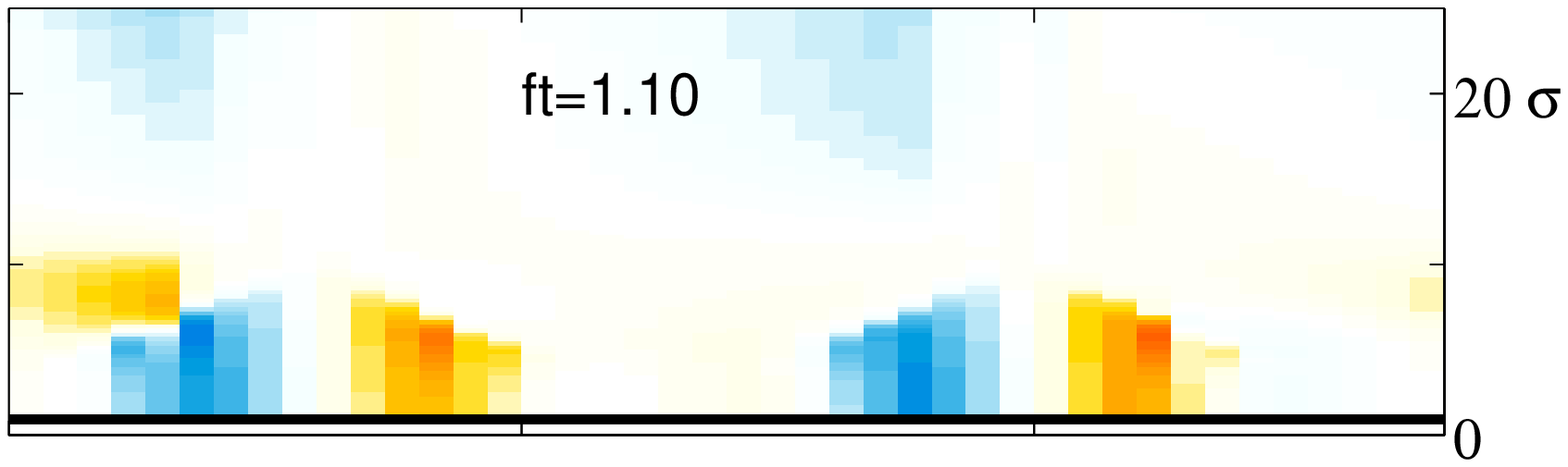}}}}\\
\vspace{-0.45cm}%
\subfloat{\scalebox{0.45}{{\includegraphics{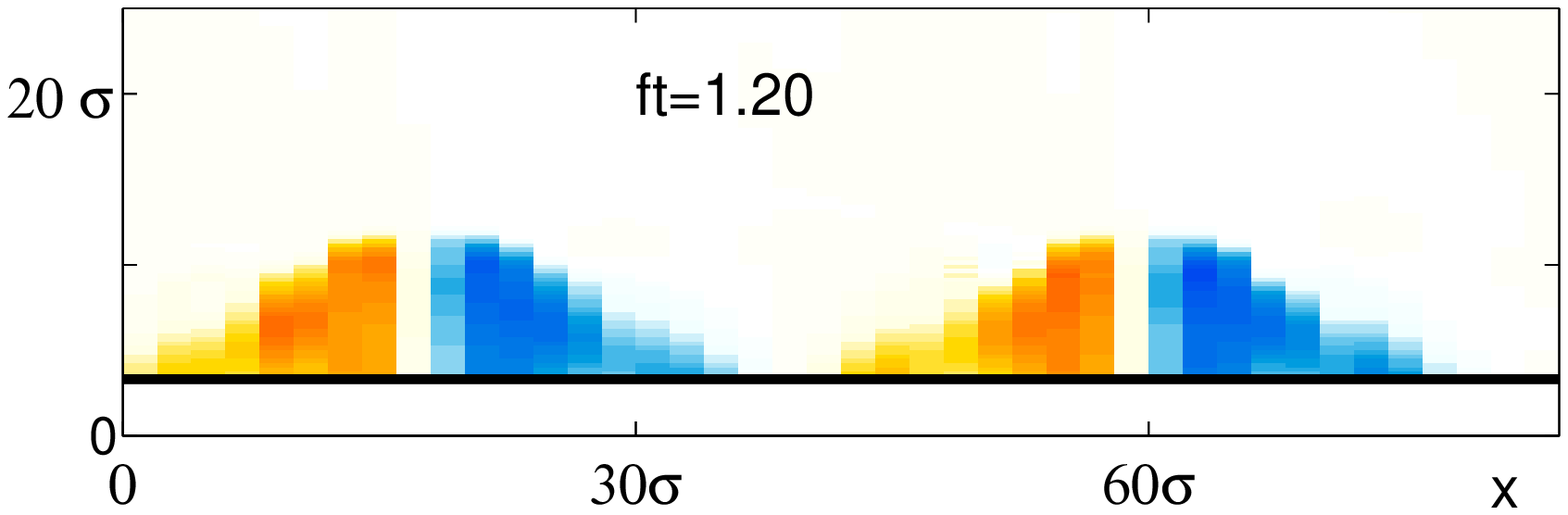}}}}
\subfloat{\scalebox{0.45}{{\includegraphics{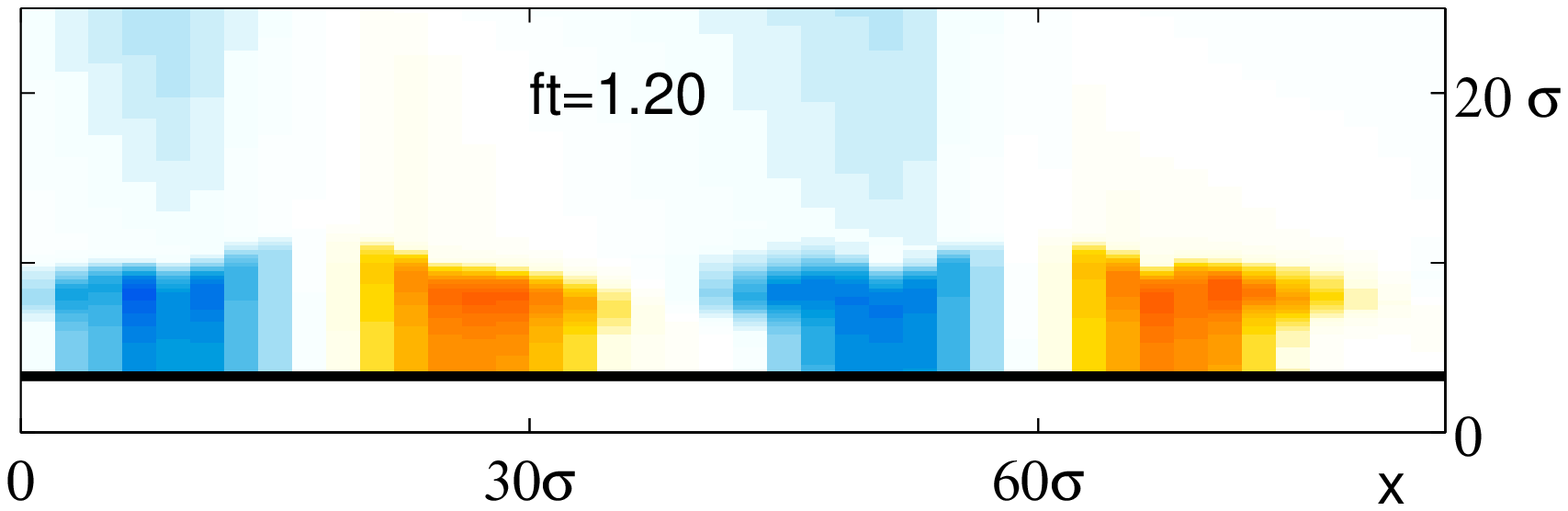}}}}\\
\hspace{0.5cm}
\subfloat{\scalebox{0.45}{{\includegraphics{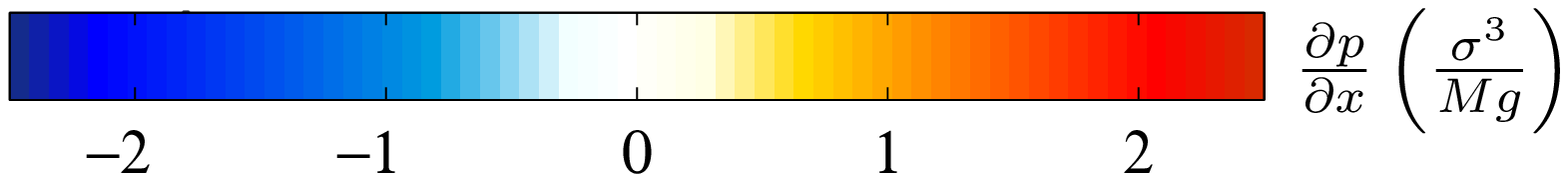}}}}
\hspace{0.2cm}
\subfloat{\scalebox{0.45}{{\includegraphics{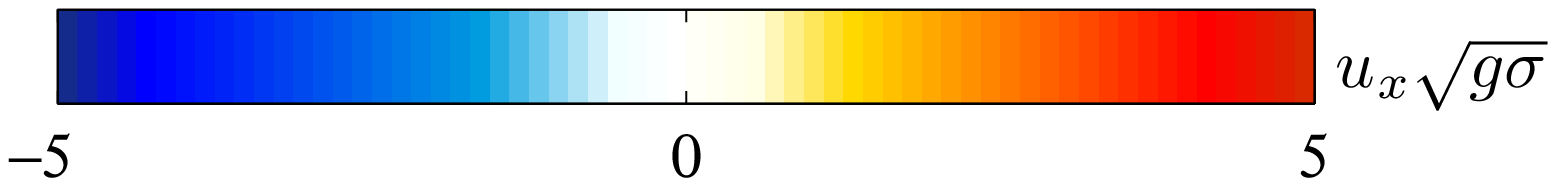}}}}\\
\caption{\label{sloshing} 
A side view of a layer of grains, 
showing 
dimensionless horizontal pressure gradient 
$\frac{\partial p}{\partial x}\left(\frac{\sigma^3}{M g}\right)$ (left column)
and
dimensionless horizontal velocity 
$u_x \sqrt{g \sigma}$ (right column) 
at a slice
$y=5\sigma$ parallel to the $x$-$z$ plane at various times $ft$.    
Although the full cell has a length
$168\sigma$ in the $x$-direction and a height $160\sigma$ in the 
$z$-direction, this figure only displays the horizontal range
$0\leq x\leq84\sigma$ and the vertical range $0\leq z\leq 25\sigma$
to show a closer view of the collision of the shock with the plate.
In the left column, 
pressure decreasing left to right ($\frac{\partial p}{\partial x}<0$)
is shown in shades of blue, 
$\frac{\partial p}{\partial x}=0$ is white, and
pressure increasing left to right ($\frac{\partial p}{\partial x}>0$)
is shown in shades of red.
In the right column, flow towards the left ($u_x<0$)
is shown in shades of blue, 
zero horizontal velocity ($\nu=0$) is white,
and flow towards the right ($u_x>0$) is shown in shades of red.
In both columns, the plate is shown as a thick, horizontal black line.
}
\end{figure*}

In order to examine the sloshing motion of the layer, 
we also
show 
the dimensionless $x$-component of the pressure gradient 
$\frac{\partial p}{\partial x}\left(\frac{\sigma^3}{M g}\right)$ (left column)
and the dimensionless horizontal velocity $u_x \sqrt{g \sigma}$ (right column)
at the these same times and locations in Fig.~\ref{sloshing}.
We now
proceed to examine the horizontal variation in flow at these
five times as the layer collides with the plate.

\subsubsection{$ft\approx0.8:$ The layer is off the plate.}

At $ft=0.8$, the layer is mostly off the plate, and peaks and valleys
are clearly visible in the left column of Fig.~\ref{shocks}.  
In the right column of Fig.~\ref{shocks}, we show the dimensionless
pressure $p\left(\frac{\sigma^2}{M g}\right)$.  
At this time, the layer is more dilute than it will be at later times
in the cycle, and
the pressure is uniformly low throughout the
layer.  There are no strong pressure gradients visible, as 
can be seen 
in the left column of Fig.~\ref{sloshing}, 
where we show the dimensionless
$x$-component of the pressure gradient
$\frac{\partial p}{\partial x}\left(\frac{\sigma^3}{M g}\right)$.

In the right column of Fig.~\ref{sloshing}, we can see that $u_x>0$
(the flow is moving towards the right) 
for $0\lesssim x \lesssim 16\sigma$; there is a small range of nearly
zero horizontal flow rate near $16\sigma\lesssim x \lesssim 20\sigma$; 
then  $u_x>0$ (flow is moving towards the left) for 
$16\sigma \lesssim x \lesssim 34 \sigma$.  Since
the peak of the pattern is located at
$16\sigma\lesssim x\lesssim 20\sigma$ at this time
({\it cf} Fig.~\ref{shocks}), grains are flowing
towards the peak from both the left and the right.  Therefore,
the flow is still removing material from the valleys and 
moving it towards
the peak; the difference between peaks and valleys is 
growing at this time.

\subsubsection{$ft\approx0.9:$ The layer begins to contact the plate.}

At $ft=0.9$, the bottom of the layer has just begun to contact
the plate at the horizontal location of the two peaks,  
but the material between the peaks has not yet contacted the plate
(Fig.~\ref{shocks}).  
The layer looks similar to its appearance
at $ft=0.8$, except that in a very small region near the plate, 
the pressure has begun to slightly increase where the layer is starting
to contact the plate.

This pressure increase at the two points of contact yields
a horizontal pressure gradient with maximum pressure near the peak
location.  
For instance, the reddish color found near the plate for 
$12\sigma\lesssim x \lesssim 18 \sigma$ in the left
column of Fig.~\ref{sloshing} at this time indicates that the pressure is
increasing from left to right; the bluish color for
$18\sigma\lesssim x \lesssim 26 \sigma$ in the same picture
indicates that the pressure decreases from left to right.

The picture in the right column for this time shows that the flow
is still moving towards the peak through the bulk of the layer at this
time.  However, the pressure gradient near the plate
tends to drive flow in the opposite direction: from high pressure (near
the peak) into low pressure (near the valley).  Therefore, the horizontal
flow velocity near the plate has begun to slow (leaving a white space
near the plate where the pressure gradient is strongest) and is even
beginning to reverse direction (note the
light blue near the plate at the location of the second peak).

\subsubsection{$ft \approx 1.0:$ A shock forms.}

By $ft=1.0$, more of the material is colliding
with the plate, 
and the pressure has noticeably increased
at the two main contact points between the layer and the plate
(Fig.~\ref{shocks}).
This increase in layer density and pressure near the plate
marks the formation of a shock ({\it cf} Fig.~\ref{onedshocks}).

As discussed in Sec.~\ref{subsection-shocks},
these shocks are not uniform horizontally; larger
layer depth near the peaks leads to larger pressure
at these locations.
Thus, shock formation leads to pressure maxima
near the plate at the horizontal location of the pattern peaks
with strong pressure gradients 
as seen in the left column of Fig.~\ref{sloshing}.

These pressure gradients drive the flow in the direction
opposite the pressure gradient; thus the direction of flow near the
plate at this time is clearly reversing (see the 
right column of Fig.~\ref{sloshing}).  For much of the layer,
the flow is still away from the valleys and towards the peaks.
Near the plate in the region of strong pressure gradients, however, the
flow has reversed itself and is now flowing away from the peaks and
towards the valleys.
This separation of the flow into two distinct domains represents the
formation of a shock near the plate.

\subsubsection{$ft \approx 1.1:$ The shock propagates.}

This shock then travels upward through the layer, 
away from the plate.
By $ft=1.1$, this shock 
is well-developed and has propagated through most of the layer, 
separating a high- density
and pressure region near the plate from a low- density and pressure
region above the shock front  
(Fig.~\ref{shocks}).

The pressure gradient is quite strong
near the plate (Fig.~\ref{sloshing})
and most of the layer is behind the shock.  The
flow behind the shock has reversed direction, while the 
material ahead of the shock has not yet changed direction.
There is still some
material in the peaks that is still falling towards the plate ahead
of the shock ({\it cf} Fig.~\ref{shocks}), but its horizontal 
velocity is small, and the pressure
is relatively low in this region (see Fig.~\ref{sloshing}).

\subsubsection{$ft \approx 1.2:$ The flow has reversed.}

By $ft=1.2$, nearly the entire layer is behind the shock and
the layer has noticeably flattened (Fig.~\ref{shocks}), 
although peaks and valleys are still visible.  
As the shock
propagates into the low-density region above the layer, the layer
cools behind the shock due to inelastic collisions.
This causes the pressure to decrease slightly at $ft=1.2$ 
as compared to $ft=1.1$.

The pressure gradient is still driving
the flow away from the peaks, although this gradient
is not as strong as in at $ft=1.1$ 
(Fig.~\ref{sloshing}).  
At this time, nearly the entire layer has reversed direction
and is now flowing away from the peaks and towards the valleys.

As this flow continues, the layer becomes flat, and then develops
peaks where there were previously valleys, and vice versa.  
Thus, the horizontal pressure gradients
created by the nonuniform shock front drive the 
flow to reverse itself.  It is this reversal that creates
the sloshing motion visible in the subharmonic oscillation.

\section{Frequency Dependence of Shocks and Standing Waves}\rm\label{section-frequency}

In Sec.~\ref{section-dynamics}, we studied the interaction between
shocks and patterns in layers oscillated with a particular 
non-dimensional frequency $f^*=0.25$.
In this section, we investigate frequency dependence by  
holding dimensionless accelerational amplitude 
$\Gamma=2.2$ constant, 
while varying dimensionless frequency $f^*$.    
Experiments have shown that wavelength $\lambda$ 
depends on the frequency of oscillation
\cite{melo1993, umbanhowar2000}.  For a range of layer depths
and oscillation frequencies, experimental data for
frictional particles near pattern onset were fit by the
function $\lambda^{*}=1.0+1.1f^{*-1.32\pm0.03}$, where
$\lambda^{*}=\lambda/H$ \cite{umbanhowar2000}.
We examine the correlation between changes
in shock properties with changes in pattern wavelength
throughout this frequency range.
For each frequency, we 
start from an artificial flat layer, then simulate 250 cycles of the
plate
to allow the layer to reach an oscillatory state.  Data is taken
from the next six cycles of the plate.

\subsection{Maximum Mach number varies inversely with driving frequency}

\begin{figure}[htb]
\scalebox{0.5}{\includegraphics{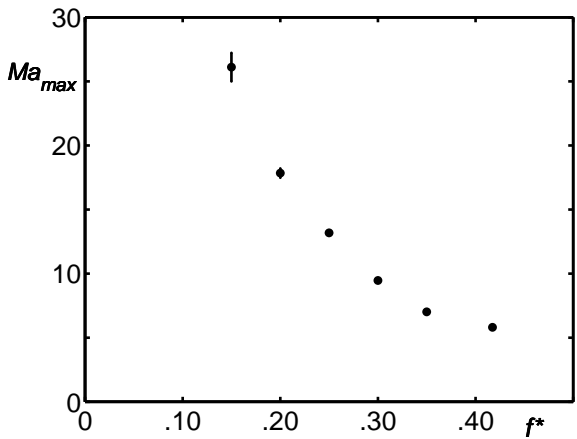}}
\caption{\label{machf} The maximum Mach number $Ma_{\rm max}$ of the
layer with respect to the plate found at any time during an
oscillation cycle as a function of the dimensionless driving frequency
$f^{*}$.  We calculate $Ma_{\rm max}$ for each of six oscillation cycles;
points are the average of the six cycles and error bars represent
 the standard deviation.}
\end{figure}

Shock properties depend on the
Mach number of the layer with respect to 
the plate during collision.
In Fig.~\ref{machf}, we plot the maximum Mach number $Ma_{\rm max}$
of the layer with respect to the plate as a function of dimensionless
driving frequency $f^{*}$.  To calculate $Ma_{\rm max}$, we find
the Mach number $Ma\left({\bf x},t\right)$
at each location ${\bf x}$ in the cell
during an oscillation cycle.  To ensure that we 
are looking at the Mach number of material in the layer, rather than
in the low-density region above the layer, at each time $t$, we find the 
highest Mach number corresponding to at least $1 \%$ of the total 
mass of the layer.
We then define $Ma_{\rm max}$ as the maximum Mach number 
found at any time during a cycle.

Note that $Ma_{\rm max}$ monotonically decreases as $f^*$ increases 
(Fig.~\ref{machf}).
For fixed $\Gamma=2.20$, lower oscillation frequency
corresponds to a higher maximum plate velocity
and higher oscillation 
amplitude.  Layers with lower $f^{*}$
will generally reach a higher maximum height 
when they leave the plate,
and will have a larger speed relative to the plate when 
they collide later in the cycle.  Although
the relative Mach number between the layer and the plate depends
on the speed of sound in the layer as well as the time during 
the cycle at which the collision takes place,
higher frequencies correspond to lower $Ma$ at collision
throughout this range.

\subsection{Higher Mach number at collision produces stronger shocks}

\begin{figure}[htb]
\subfloat{\label{pzmach}\scalebox{0.5}{\includegraphics{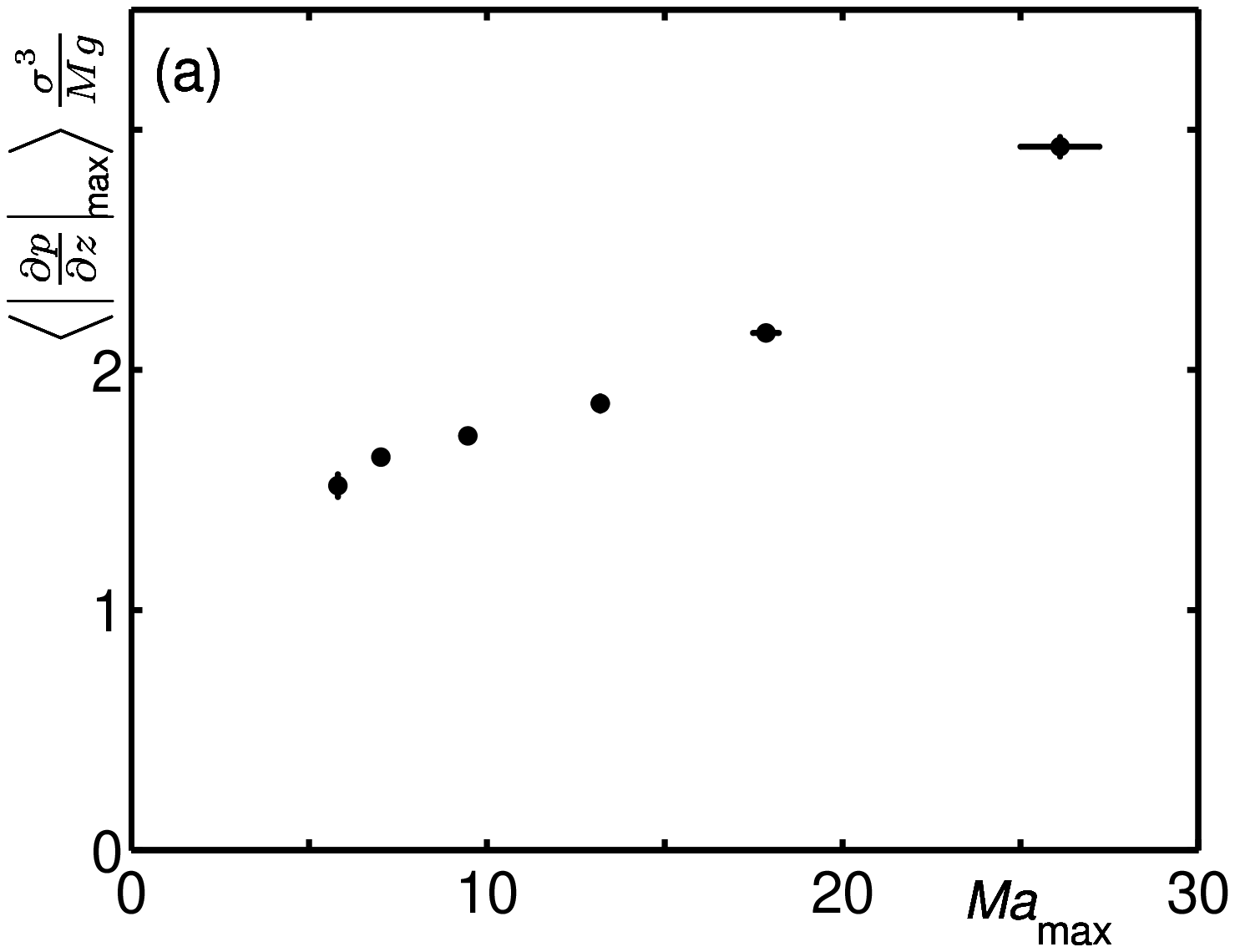}}}\\
\subfloat{\label{pxpz}\scalebox{0.5}{\includegraphics{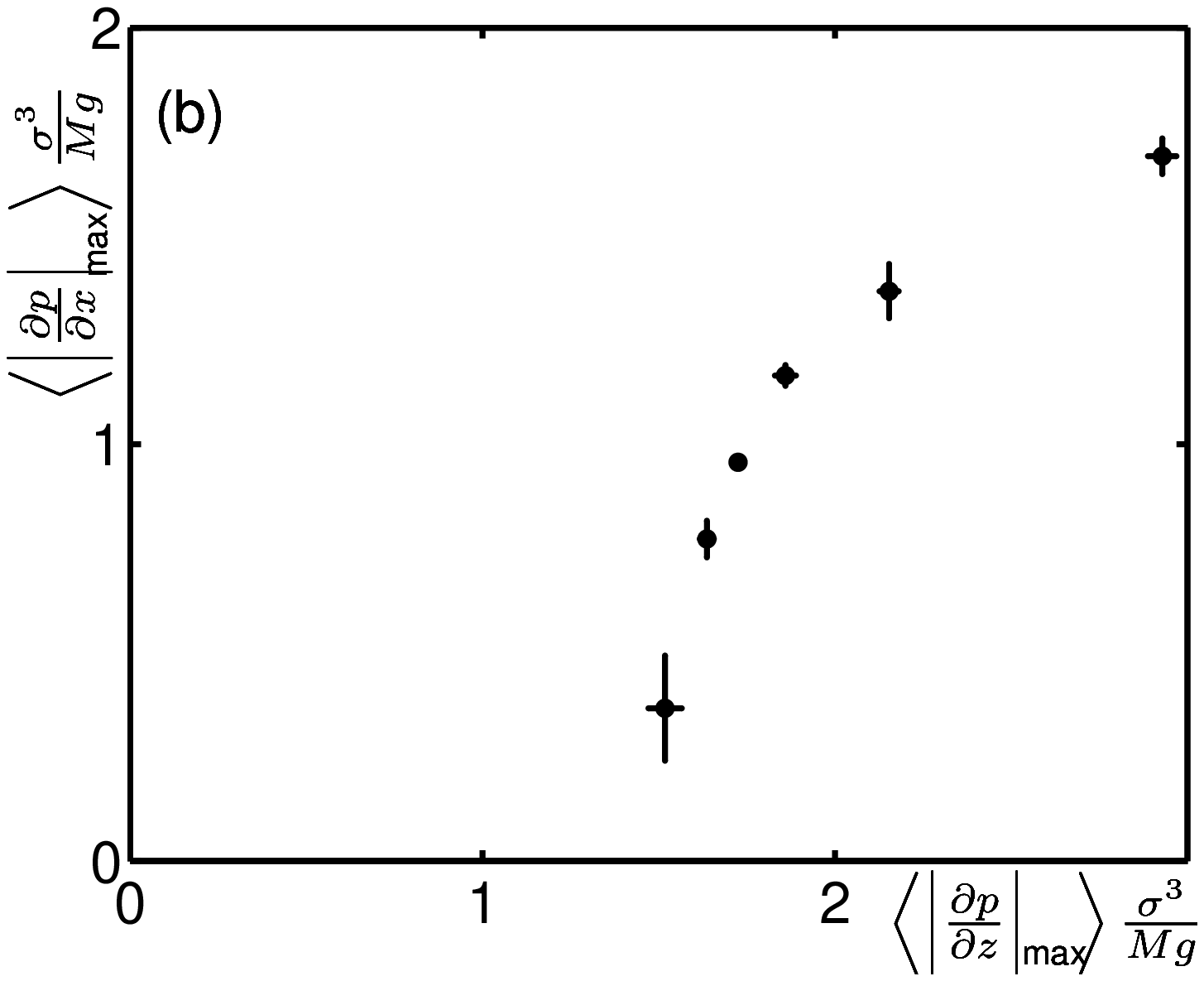}}}
\caption{\label{machpressure} 
The dimensionless maximum $z$-component (a) of the pressure 
gradient found anywhere in the cell averaged over all times in the cycle
$\left<\left|\frac{\partial p}{\partial z}\right|_{\rm max}\right>\frac{\sigma^3}{M g}$ 
as a function of 
the maximum Mach number $Ma_{\rm max}$ of the
layer with respect to the plate found at any time during an
oscillation cycle, and the
maximum $x$-component (b) of the pressure 
gradient found anywhere in the cell averaged over all times in the cycle
$\left<\left|\frac{\partial p}{\partial x}\right|_{\rm max}\right>\frac{\sigma^3}{M g}$ 
as a function of
the maximum $z$-component of the pressure 
gradient found anywhere in the cell averaged over all times in the cycle
$\left<\left|\frac{\partial p}{\partial z}\right|_{\rm max}\right>\frac{\sigma^3}{M g}$.
We calculate values for each of six oscillation cycles;
points are the average of the six cycles and error bars represent
 the standard deviation.}
\end{figure}

Pressure decreases rapidly when moving
from the region behind a shock, across the shock front,
to the undisturbed region
ahead of the shock.
We examine this pressure change by calculating the 
magnitude of the $z$-component of the pressure
gradient 
$\left| \frac{\partial p}{\partial z}\left({\bf x},t\right)\right|$ 
at each
time $t$ and location in the cell ${\bf x}$.  We then find
$\left|\frac{\partial p}{\partial z}\right|_{\rm max}\left(t\right)$ 
as the highest value
corresponding to at least $1 \%$ of the total mass of the layer.
Finally, we calculate the time average of these layers 
$\left<\left|\frac{\partial p}{\partial z}\right|_{\rm max}\right>$,
nondimensionalize by a factor of $\frac{\sigma^3}{M g}$, 
and plot this
as a function of $Ma_{\rm max}$ in Fig.~\ref{pzmach}.  
As 
the maximum Mach number $Ma_{\rm max}$ increases,
the time-averaged maximum vertical pressure gradient 
increases monotonically.
In other words, the higher the relative Mach number of the layer
with respect to the plate, the sharper the pressure drop across the
shock.  
Note that hydrostatic pressure
would cause pressure to vary with depth even in static layers,
consistent with the non-zero intercept of Fig.~\ref{pzmach}.

As discussed in Sec.~\ref{section-dynamics}, although the shock
is produced by collision with a vertically oscillating plate, 
horizontal pressure variation develops as a result of the peaks 
and valleys in the layer.  Shocks with 
stronger vertical pressure gradients correspond to 
stronger horizontal pressure gradients as well (Fig.~\ref{pxpz}).
A flat layer with no patterns would be expected to have pressure
variation with depth but not with horizontal location. This is 
consistent with the apparent non-zero intercept in Fig.~\ref{pxpz}.

\subsection{Horizontal pressure gradients drive horizontal velocity}

\begin{figure}[htb]
\scalebox{0.5}{\includegraphics{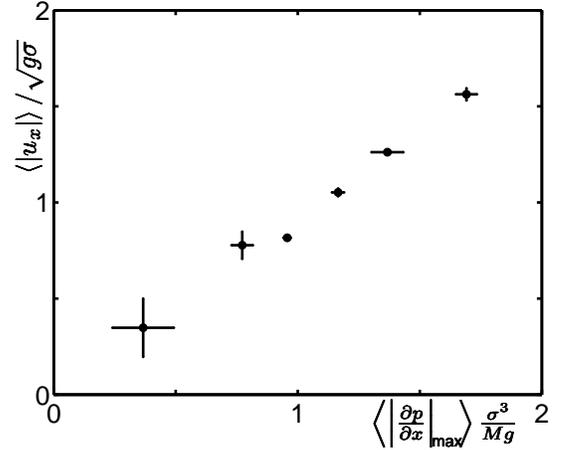}}
\caption{\label{upx} 
The nondimensional average flow
speed in the $x$-direction 
$\left<\left|u_x\right|\right>/\sqrt{g \sigma}$
as a function of the 
dimensionless maximum $x$-component of the pressure 
gradient found anywhere in the cell averaged 
over all times in the cycle
$\left<\left|\frac{\partial p}{\partial x}\right|_{\rm max}\right>\frac{\sigma^3}{M g}$.
Points are calculated as the average of six oscillation 
cycles of the plate, and error bars represent
 the standard deviation of these six cycles.}
\end{figure}

As strong pressure gradients develop in the layer, there
will be a tendency for the material to flow from high
pressure to low pressure; i.e. the direction of 
flow velocity will be
opposite the pressure gradient ({\it cf} Fig.~\ref{sloshing}). 
We calculate the average flow speed in the $x$-direction
at all times and all locations in the cycle 
$\left<\left|u_x\right|\right>$.
Figure~\ref{upx} shows that the nondimensionalized average flow
speed in the $x-$direction 
$\left<\left|u_x\right|\right>/\sqrt{g \sigma}$ increases
monotonically as the maximum dimensionless 
$x$-component of the pressure 
gradient found anywhere in the cell averaged 
over all times in the cycle
$\left<\left|\frac{\partial p}{\partial x}\right|_{\rm max}\right>\frac{\sigma^3}{M g}$ increases.
Thus, the stronger the maximum 
horizontal pressure gradients produced
across the shock are, the faster the average horizontal flow
speed becomes.

\subsection{Horizontal velocity produces standing waves}

\begin{figure}[htb]
\scalebox{0.5}{\includegraphics{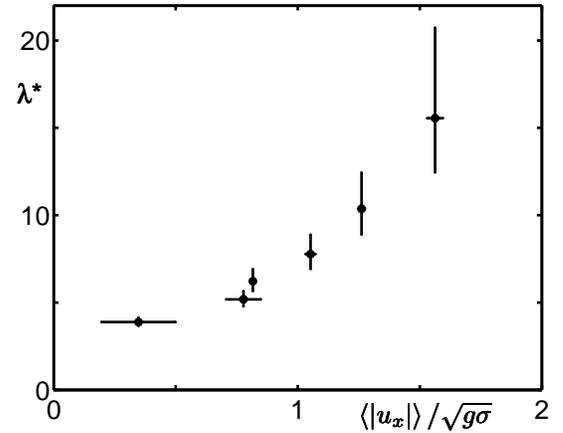}}
\caption{\label{lambdau} 
Nondimensional wavelength
$\lambda^{*}=\lambda/H$ 
as a function of nondimensional average flow
speed in the $x$-direction 
$\left<\left|u_x\right|\right>/\sqrt{g \sigma}$.
Points are calculated as the average of six oscillation 
cycles of the plate.  The error bars for
$\lambda^{*}=\lambda/H$ are
calculated exclusively from discretization due to periodic boundary
conditions in a finite-size box. 
The error bars for 
$\left<\left|u_x\right|\right>/\sqrt{g \sigma}$ represent
 the standard deviation of these six cycles.
 }
\end{figure}

For the dimensionless frequency $f^{*}=0.25$ examined in 
Sec.~\ref{section-dynamics}, four wavelengths fit in 
a box of size
$164 \sigma$ in the $x$-direction, 
yielding a wavelength
of $41\sigma$ (Fig.~\ref{patterns}).  
We calculate the dimensionless wavelength $\lambda^{*}=\lambda/H$,
and plot it as a function of
nondimensional average flow
speed in the $x$-direction 
$\left<\left|u_x\right|\right>/\sqrt{g \sigma}$ for each
frequency in our range (Fig.~\ref{lambdau}).
Due to the periodic boundary
conditions and finite size of the box, 
an integer number of waves must fit in the box.  
This finite size
effect of quantized wavelength yields inherent 
uncertainty in the wavelength
that would be selected in an infinite box.

Faster flow in the $x$-direction corresponds to more horizontal
motion of the particles throughout an oscillation cycle.
Therefore, the wavelength
(and therefore the distance between peaks)
increases monotonically as the average horizontal flow speed
increases.
(see Fig.~\ref{lambdau}).

\subsection{Dispersion relation}

Summarizing the results from this section thus far, we find
that higher frequencies of oscillation yield smaller
Mach numbers between the layer and the plate during collision
(Fig.~\ref{machf}).  Larger values of the Mach number
produce stronger shocks, as reflected in
larger vertical and horizontal changes in pressure
across the shock front (Fig.~\ref{machpressure}).
Larger pressure gradients produce faster horizontal flow
(Fig.~\ref{upx}) which in turn corresponds
to larger horizontal distances between peaks of the pattern,
and thus a larger wavelength (Fig.~\ref{lambdau}).

\begin{figure}[htb]
\scalebox{0.5}{\includegraphics{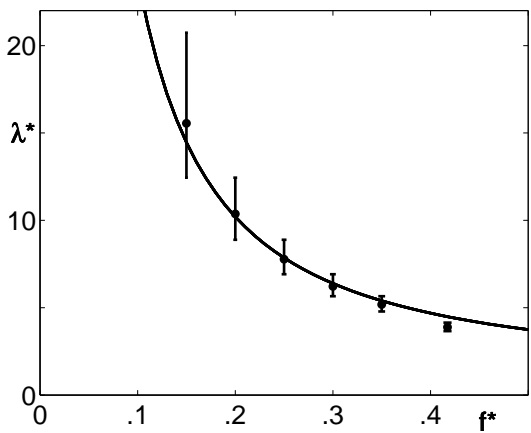}}
\caption{\label{dispersion} Dimensionless wavelength $\lambda^{*}$
as a function of dimensionless frequency $f^{*}$. 
Data from our simulations are shown as points, with
error bars 
calculated exclusively from discretization due to periodic boundary
conditions in a finite-size box. The dominant wavelength
found in our simulations fits quite well the
dispersion relation
$\lambda^{*}=1.0+1.1f^{*-1.32\pm0.03}$ (solid line) found as a fit
to previous experimental data
\cite{umbanhowar2000}. }
\end{figure}

Given this chain of reasoning, it follows that as
the Mach number of the layer with respect to the plate
increases, wavelength of the pattern increases monotonically
throughout the range.  As Mach number increases with 
decreasing shaking frequency, 
wavelength should decrease
monotonically as the oscillation frequency increases.  
Figure~\ref{dispersion} shows that
this is indeed the case throughout the range of
frequencies we have simulated.  Additionally, the
wavelengths found in our simulation throughout this range
are consistent with the dispersion relation
previously found by fit to experimental data
\cite{umbanhowar2000}.

\section{Conclusions}\rm\label{section-conclusions}

We have simulated
vertically oscillated layers of granular media
by numerically solving a proposed set of granular hydrodynamic
equations to Navier-Stokes order.  
We have shown that these continuum simulations
can describe important aspects of shocks and patterns
in granular materials.  In our simulations, layers
spontaneously form subharmonic standing-wave patterns in 
which the wavelength $\lambda$ depends on the frequency $f$
of the oscillation of the plate.  The wavelengths
of the patterns formed
in our simulation agree with the dispersion relations
found in previous experiments.

As these standing waves oscillate subharmonically,
shocks are formed with each collision of the layer with the plate.
We have analyzed these shocks and found that 
variation in the layer depth produces a shock front that
is not uniform horizontally.
This horizontal variation
in the shock front produces horizontal components of flow 
velocity, causing grains to move from high pressure
to low pressure.  
Therefore,
despite the fact that the plate oscillates vertically, 
the shocks produced during collision between the
layer and the plate drive the
horizontal sloshing motion that 
characterizes the standing-wave patterns.

We have simulated shocks in layers oscillated at various 
frequencies, and have shown that the strength of the shock
varies depending on the oscillation frequency.
We have also established that there is a relationship
between pattern wavelength and 
the Mach number of the layer with respect to the
plate at the time of collision.
Specifically, the horizontal and vertical components
of the pressure gradient both increase monotonically as the
Mach number increases.
The horizontal flow speed increases monotonically with 
the strength of these pressure gradients, leading to 
longer wavelengths.

Therefore, throughout the range we studied, the wavelength 
of the pattern increases monotonically with the maximum
$Ma$ of the layer with respect to the plate.
This analysis indicates that shocks play a significant role
in the dynamics of the standing-wave patterns formed
in these oscillating layers.  

\begin{acknowledgments}
This research was supported by an award from Re-
search Corporation for Science Advancement.
\end{acknowledgments}


\end{document}